\documentclass[11pt,hyper,a4paper]{article}
\usepackage{jheppub}
\usepackage{amsmath,amssymb,amsfonts,amscd,bm, amsthm, slashed, mathrsfs, bm}
\usepackage{graphicx, wrapfig}
\usepackage{multirow}
\usepackage{verbatim}
\usepackage[title]{appendix}
\usepackage{fancybox}
\usepackage[utf8]{inputenc}
\usepackage{arydshln}

\usepackage{url}
\usepackage{float}

\usepackage[normalem]{ulem}
\usepackage{graphicx}

\usepackage{color}
\usepackage[usenames,dvipsnames,svgnames,table]{xcolor}
\definecolor{darkblue}{cmyk}{0.9,0.9,0,0}
\hypersetup{colorlinks=false, linkcolor=darkblue,linkcolor=darkblue, citecolor=darkblue}

\usepackage{multirow}
\usepackage{verbatim}

\usepackage{bbm}
\usepackage{comment}
\usepackage{enumitem}
\usepackage{float}
\usepackage{empheq}
\usepackage[usenames,dvipsnames,svgnames,table]{xcolor}
\usepackage{enumerate}

\title{BPS Invariants for a Knot in Seifert Manifolds}

\author{Hee-Joong Chung}

\affiliation{Department of Science Education, Jeju National University, Jeju, 63243, Republic of Korea}

\abstract{We calculate homological blocks for a knot in Seifert manifolds when the gauge group is $SU(N)$.
We obtain the homological blocks with a given representation of the gauge group from the expectation value of the Wilson loop operator by analytically continuing the Chern-Simons level.
We also obtain homological blocks with the analytically continued level and representation for a knot in the Seifert integer homology spheres.}

\begin{document}

\maketitle

\section{Introduction}
\label{sec:intro}
Knot invariants have been studied actively both in mathematics and physics due to their importance.
Knot polynomials can be understood as the Wilson loop expectation value in Chern-Simons theory \cite{Witten-Jones} and since the work of Witten various physical aspects of knot polynomials have been discussed in literature.

Recently, there have been important developments in knot polynomials, which are about the analytic continuation of them \cite{Gukov-Pei-Putrov-Vafa,Gukov-Manolescu}.
Chern-Simons partition function for a knot $K$ in a non-trivial 3-manifold $M_3$ in general is not a $q$-series with integer powers and integer coefficients.
But it was conjectured in \cite{Gukov-Pei-Putrov-Vafa} that the Wilson loop expectation value with a representation $\mathcal{R}$ of the gauge group $G$ can be decomposed into $q$-series with integer powers and integer coefficients, which is called the homological block $\widehat{Z}(M_3,K;q, \mathcal{R})$, as in the case of closed 3-manifolds \cite{Gukov-Putrov-Vafa, Gukov-Pei-Putrov-Vafa}.
For $G=SU(2)$, the conjecture states that
\begin{align}
Z_{SU(2)}(M_3, K; k, \mathcal{R}) \simeq \sum_{a,b \in \text{Tor} H_1(M_3,\mathbb{Z})/\mathbb{Z}_2} e^{2\pi i CS_a} S_{ab} \widehat{Z}_b(M_3,K;q, \mathcal{R}) \Big|_{q \searrow e^{\frac{2\pi i}{k}}}	\,	.	\label{decompz}
\end{align}
Here, $CS_a$ is a Chern-Simons invariant of abelian flat connection $a$, and $S_{ab}$ is the $S$-transform that satisfies $S^2=I$ where $I$ is an identity.
$q \searrow e^{\frac{2\pi i}{k}}$ means that $\tau$ of $q=e^{2\pi i \tau}$ approaches to $1/k$, $k \in \mathbb{Z}$ from the upper half plane of $\tau$.
The decomposition \eqref{decompz} was checked for the lens space in \cite{Gukov-Pei-Putrov-Vafa}.
These homological blocks can be regarded as the invariants that provide an analytic continuation of the Chern-Simons partition function for a knot in non-trivial 3-manifolds with a given representation $\mathcal{R}$, which are expected to be categorified.

Meanwhile, it was conjectured in \cite{Gukov-Manolescu} that there is a two-variable knot invariant that is obtained from the Borel resummation of the Melvin-Morton-Rozansky expansion \cite{Melvin-Morton, Rozansky-higherMM, Bar-Natan-Garoufalidis} of the colored Jones polynomial.
It is often denoted as $F_K(x,q)$ in literature and can be regarded as an analytic continuation of the colored Jones polynomial with both the Chern-Simons level $k$ and the representation $\mathcal{R}$ analytically continued.
A number of aspects of $F_K(x,q)$ including calculations and generalizations have been studied, for example, in \cite{Park-ZF,Park-largeR,EGGKPS-largeN,Kucharski:2020rsp,Chung-resurg,Park-inverted,EGGKPS-branches}.	\\

In the context of M-theory, a knot that gives the homological block $\widehat{Z}(M_3,K;q,\mathcal{R})$ arises from additional M2 branes
\begin{align}
\begin{tabular}{r c c c c c c}
\text{space-time}		&	&	$\mathbb{R}$ 	&$\times$ 	&$TN$ 	&$\times$ 	&$T^* M_3$	\\
\text{M5 branes}		&	&	$\mathbb{R}$ 	&$\times$ 	&$D^2$ 	&$\times$ 	&$M_3$	\\
\text{M2 branes}		&	&	$\mathbb{R}$ 	&$\times$ 	&$pt$ 	&$\times$ 	&$T^*K$	\\
\end{tabular}
\label{mconf1}
\end{align}
where $D^2$ is a disc, $TN$ is Taub-NUT space, $D^2 \subset TN$, and $pt$ is a point at the center of $D^2$.
Or a knot can also be realized from additional M5 branes that are supported on $\mathbb{R} \times L_K$ where $L_K$ is a conormal bundle of $K$ such that $L_K \subset T^* M_3$ and $L_K \cap M_3 = K$ and on the contangent space to $D^2$ at the point $pt$ \cite{Ooguri-Vafa}.
Both of them lead to a Wilson loop in Chern-Simons theory.\footnote{These were also studied in the context of the 3d-3d correspondence, for example, in \cite{Gang:2015wya}.}
In these systems, $U(1)_R$ symmetry and $U(1)_{q}$ symmetry that is given by a linear combination of the rotational symmetry on $D^2$ and $U(1)_R$ symmetry are preserved, and they provide two gradings in the homological invariants that lead to homological blocks.
For a Seifert manifold $M_3$, there is an additional $U(1)_\beta$ symmetry that arises due to the existence of the semi-free $U(1)$ action on the Seifert manifold, which lead to another extra grading in the homological invariants.
In the context of the 3d-3d correspondence \cite{Dimofte-Gaiotto-Gukov, Chung-Dimofte-Gukov-Sulkowski}, the homological block is given by the $D^2 \times_q S^1$ partition function or the half index of the 3d $\mathcal{N}=2$ theory $T[M_3]$ with a loop operator where the boundary condition is determined by the abelian flat connection $b$ \cite{Gukov-Putrov-Vafa, Gukov-Pei-Putrov-Vafa}.

A knot that leads to the homological block $F(M_3,K; x,q)$\footnote{In our notation, $F(S^3,K; x,q)=F_K(x,q)$.} arise from different additional M5 branes 
\begin{align}
\begin{tabular}{r c c c c c c}
\text{space-time}		&	&	$\mathbb{R}$ 	&$\times$ 	&$TN$ 	&$\times$ 	&$T^* M_3$	\\
\text{M5 branes}		&	&	$\mathbb{R}$ 	&$\times$ 	&$D^2$ 	&$\times$ 	&$M_3$	\\
\text{M5$^{\prime}$ branes}		&	&	$\mathbb{R}$ 	&$\times$ 	&$D^2$ 	&$\times$ 	&$L_K$	\\
\end{tabular}	\,	.
\label{mconf2}
\end{align}
Differently from the previous additional M5 branes in \eqref{mconf1}, M5$^{\prime}$ branes in \eqref{mconf2} is supported on $D^2$.
In the context of the 3d-3d correspondence, this M5$^{\prime}$ branes changes the theory $T[M_3]$ to $T[M_3\backslash K]$ while the additional M5 branes or M2 branes in \eqref{mconf1} give a loop operator in $T[M_3]$.
The homological block $F(M_3,K; x,q)$ is given by the half index of $T[M_3\backslash K]$ with the boundary condition determined by the abelian flat connection.	\\

In this paper, we calculate homological blocks $\widehat{Z}(M_3,K;q,\mathcal{R})$ and $F(M_3,K; x,q)$ for a knot $K$ in an infinite family of Seifert manifolds $M_3$ when the gauge group is $SU(N)$.
The knot $K$ is supported on $S^1$ fiber of the Seifert manifold and it is called the Seifert knot.
From the expression of the Chern-Simons partition function with the Wilson loop supported on a knot in Seifert manifold, which was obtained in \cite{Beasley-Wilson, Blau-Thompson-Seifert}, by analytically continuing the Chern-Simons level $k$ only or both the level $k$ and the representation $\mathcal{R}$, we obtain homological blocks.
We see that the structure or the decomposition of Wilson loop expectation values in terms of homological blocks also arises for the analytically continued $k$ as in the case of closed 3-manifolds.
However, when $\mathcal{R}$ is also analytically continued and for a Seifert rational homology sphere, we don't obtain homological blocks while having the structure as in the case of closed 3-manifolds except for the case of a knot in Seifert integer homology spheres.

In section 2, we consider the case $G=SU(2)$, which is simpler and shows some essential features.
In section 3, we consider the higher rank case of $G=SU(N)$.
In appendix \ref{sec:lens}, we consider homological blocks for a knot in a lens space and compare them with the results in \cite{Gukov-Pei-Putrov-Vafa}.
In appendix \ref{sec:rmk}, we discuss a few more about $F(M_3,K; x,q)$ for rational Seifert homology spheres $M_3$.	\\

\noindent \textit{Note added:} While preparing the manuscript, we found that \cite{CCFFGHP} appeared, which overlaps with parts of results in this paper.

\section{Homological blocks for $G=SU(2)$}
\label{sec:su2}

We consider the Seifert manifold $X(P_1/Q_1, \ldots, P_F/Q_F)$ with $F$ singular fibers where the Seifert invariant $P_j$ and $Q_j$ are coprime for each $j=1,\ldots, F$ and $P_i$ and $P_j$ are pairwise coprime.
We set the Euler number of the $S^1$ bundle of the Seifert fibration to be zero.
We denote $P$ and $H$ by
\begin{align}
P = \prod_{j=1}^{F} P_j	\,	,	\quad	H = P \sum_{j=1} \frac{Q_j}{P_j} = \pm |\text{Tor}(H_1(M_3,\mathbb{Z}))|
\end{align}
and in this paper we choose $H$ to be positive.	\\

When $G=SU(2)$, the partition function for Seifert knot is given by
\begin{align}
\begin{split}
Z_{SU(2)}(k,R)=-\frac{1}{2 \sqrt{P}} e^{\frac{3\pi i}{4}} q^{- \frac{1}{4} (\theta_0 + (R^2-1)P)} \Bigg[ \frac{1}{2\pi i} \sum_{t=0}^{H-1} \int_{C_t} dy \, e^{-\frac{k}{2\pi i} \frac{H}{P}y^2 -2 k t y} \chi_R(y) \frac{\prod_{j=1}^{F} e^{y/P_j} - e^{-y/P_j}}{(e^y - e^{-y})^{F-2}} 	\\
-\sum_{m=1}^{2P-1} \text{Res} \bigg( e^{-\frac{k}{2\pi i} \frac{H}{P}y^2 -2 k t y} \frac{\chi_R(y)}{1-e^{-2ky}} \frac{\prod_{j=1}^{F} e^{y/P_j} - e^{-y/P_j}}{(e^y - e^{-y})^{F-2}}
\bigg)\bigg|_{y=\pi i m}	\\
-\sum_{t=1}^{H-1}\sum_{m=1}^{\lfloor \frac{2Pt}{H} \rfloor} \text{Res} \bigg( e^{-\frac{k}{2\pi i} \frac{H}{P}y^2 -2 k t y} \chi_R(y) \frac{\prod_{j=1}^{F} e^{y/P_j} - e^{-y/P_j}}{(e^y - e^{-y})^{F-2}}
\bigg)\bigg|_{y=-\pi i m}	
\Bigg]
\end{split}	\label{su2full}
\end{align}
where 
\begin{align}
\chi_R(x)=\frac{\sinh Rx}{\sinh x}
\end{align} 
is a character of the $R$-dimensional representation of $G=SU(2)$ \cite{Beasley-Wilson}.
The contour $C_t$ is parallel to a line from $-(1+i )\infty$ to $(1+i)\infty$ in the complex $y$-plane and passes at $y=-2\pi i \frac{P}{H}t$ on the imaginary $y$-axis.
Res denotes the residue and $\theta_0$ is given by
\begin{align}
\theta_0 = 3+  \sum_{j=1}^{F} 12 s(Q_j,P_j) - \frac{Q_j}{P_j}
\end{align}
where $s(Q_j,P_j)$ is the Dedekind symbol
\begin{align}
s(Q,P) = \frac{1}{4P} \sum_{l=1}^{P-1} \cot \Big( \frac{\pi l}{P} \Big) \cot \Big( \frac{\pi Q l}{P} \Big)	\,	
\end{align}
for $P>0$.

\subsubsection*{Resurgence}

Before calculating homological blocks from the Gaussian integral part of \eqref{su2full}, we check that the contributions from the abelian flat connections capture the contributions from non-abelian flat connections via resurgent analysis.
Some aspects of resurgence for the case of the torus knot in $S^3$, which is a special case of Seifert knot, have been discussed in \cite{Gukov-Manolescu,Chung-resurg}.
The Borel resummation of the Borel sum of the perturbative expansion around the abelian flat connection of $Z(k,R)$ is given by the Gaussian integral part of \eqref{su2full} with a contour $\gamma$ that is parallel to the imaginary axis and located at $\epsilon$ on the real axis of the complex $y$-plane, 
\begin{align}
\frac{1}{2\pi i} \sum_{t=0}^{H-1} \int_{\gamma} dy \, e^{-\frac{k}{2\pi i} \frac{H}{P}y^2 -2 k t y} \chi_R(y) \frac{\prod_{j=1}^{F} e^{y/P_j} - e^{-y/P_j}}{(e^y - e^{-y})^{F-2}}	
\label{su2int}
\end{align}
where $R$ and $k$ are analytically continued.
We can see that \eqref{su2int} captures the contributions from non-abelian flat connections by considering the Stoke phenomena.
For explicitness, we consider the case of $F=3$, and other numbers of singular fibers can be done similarly.

For the check, we evaluate the residue part of \eqref{su2full} and show that such residues can be recovered from the contributions from the poles that are swept by the change of the integration contour from $\gamma$ to $C_t$ for each $t$ as $k\rightarrow \mathbb{Z}_+$ and $R \rightarrow \mathbb{Z}_+$.
The first type of residues of \eqref{su2full} gives
\begin{align}
\begin{split}
-\sum_{m=1}^{2P-1} \text{Res} \bigg( e^{-\frac{k}{2\pi i} \frac{H}{P}y^2 -2 k t y} \frac{\chi_R(y)}{1-e^{-2ky}} \frac{\prod_{j=1}^{3} e^{y/P_j} - e^{-y/P_j}}{e^y - e^{-y}}
\bigg)\bigg|_{y=\pi i m}	\\
=
\sum_{m=1}^{2P-1} 2 i R (-1)^{Rm} H e^{-\frac{\pi i}{2} K \frac{H}{P} m^2} \frac{m}{P} \prod_{j=1}^3 \sin \frac{\pi m}{P_j}	\,	.	\label{su2res1}
\end{split}
\end{align}

Meanwhile, we take $k\rightarrow \mathbb{Z}_+$ and $R \rightarrow \mathbb{Z}_+$ in \eqref{su2int} and change the integration cycle to $C_0$ accordingly, then along the way the contour picks the poles in the negative imaginary axis of the $y$-plane.
For each $t$, the residues for such poles are
\begin{align}
\sum_{m=1}^{\infty} \text{Res} \bigg( e^{-\frac{k}{2\pi i} \frac{H}{P}y^2 -2 k t y} \chi_R(y) \frac{\prod_{j=1}^{3} e^{y/P_j} - e^{-y/P_j}}{e^y - e^{-y}}
\bigg)\bigg|_{y=-\pi i m}	\,	.
\end{align}
After the regularization, and then summing over $t=0, \ldots, H-1$, we see that this gives the same result as \eqref{su2res1}.
Then by shifting the contour to $y=-2 \pi i \frac{P}{H} t$ for each $t$, we obtain the rest of the residues in \eqref{su2full}.
Thus, the residue parts can be recovered from the contributions of the abelian flat connections to the partition function.

This resurgent analysis indicates that \eqref{su2int} would give the homological block for Seifert knot.
We first consider the case of a single abelian flat connection, $H=1$.

\subsection{The case of a single abelian flat connection}

We consider both cases that $R$ is an integer and analytically continued.

\subsubsection*{Integer $R$}

When $G=SU(2)$, the condition that $R$ is an integer can be used from the beginning of the calculation of \eqref{su2int} by taking 
\begin{align}
\chi_R(y) = \frac{e^{Ry}-e^{-Ry}}{e^y - e^{-y}} = \sum_{j=0}^{R-1} e^{(R-1-2j)y}	\,	.	\label{char}
\end{align}
Then, when $F=3$ and $H=1$, \eqref{su2int} becomes
\begin{align}
\int_{\gamma} dy \, e^{-\frac{k}{2\pi i} \frac{H}{P}y^2} \sum_{j=0}^{R-1} e^{(R-1-2j)y} \frac{\prod_{j=1}^{F} e^{y/P_j} - e^{-y/P_j}}{e^y - e^{-y}}		\,	.	\label{intr}
\end{align}
We expand the rational function of sine hyperbolic functions in \eqref{intr}
\begin{align}
\frac{\sinh (P+l)y }{\sinh Py} &= \sum_{m=0}^{M_l} (e^{(l-2mP)y} + e^{-(l-2mP)y}) - \sum_{n=0}^{\infty} \psi_{2P}^{(l-2M_l P)}(n) e^{-ny}	\label{sinh}
\end{align}
for $M_l=-1,0,1, \ldots$ such that $2M_l P < l < 2(M_l+1)P$ where
\begin{align}
\psi_{2P}^{(l)}(n) = 
\begin{cases}
\pm 1	&	\text{if } n \equiv \pm l \text{ mod } 2P	\\
0	&	\text{otherwise}
\end{cases}	\,	.
\end{align}
By using \eqref{sinh}, \eqref{intr} can be expressed as 
\begin{align}
\begin{split}
\sum_{j=0}^{R-1} \sum_{s=0}^3 \sum_{m=0}^{M_{Y_s}} (q^{\frac{P}{4}((R-1-2j)+(Y_s/P+2m))^2} + q^{\frac{P}{4}((R-1-2j)-(Y_s/P+2m))^2})	\\
+ \sum_{j=0}^{R-1} \sum_{n=0}^{\infty} \mathcal{A}_{2P}(n) q^{\frac{P}{4}((R-1-2j)-n/P)^2} \bigg|_{q \searrow e^{\frac{2\pi i}{k}}}	\,	.
\end{split}	\label{intr2}
\end{align}
where
\begin{align}
\mathcal{A}_{2P}(n) := \sum_{s=0}^3 \psi_{2P}^{(Y_s)}(n)
\end{align}
and
\begin{align}
\begin{split}
Y_0 &= P\bigg(1-\bigg(\frac{1}{P_1}+\frac{1}{P_2}+\frac{1}{P_3}\bigg)\bigg)	\,	,	\
\hspace{12mm}Y_1 = P\bigg(1-\bigg(\frac{1}{P_1}-\frac{1}{P_2}-\frac{1}{P_3}\bigg)\bigg)	\,	,	\\
Y_2 &= P\bigg(1-\bigg(-\frac{1}{P_1}+\frac{1}{P_2}-\frac{1}{P_3}\bigg)\bigg)	\,	,	\
\hspace{7.5mm}Y_3 = P\bigg(1-\bigg(-\frac{1}{P_1}-\frac{1}{P_2}+\frac{1}{P_3}\bigg)\bigg)	\,	.
\end{split}
\label{valuey}
\end{align}
Only when $(P_1,P_2,P_3)=(2,3,5)$, a term in the first line of \eqref{intr2} doesn't vanish and is given by
\begin{align}
\sum_{j=0}^{R-1} q^{\frac{30}{4}((R-1-2j)+1/30)^2} + q^{\frac{30}{4}((R-1-2j)-1/30)^2}	\,	.
\end{align}

\subsubsection*{Analytically continued $R$}

Since the character is no longer a polynomial when $R$ is analytically continued, for the calculation of the homological block with an analytically continued $R$, we calculate the expansion of $\frac{\sinh ly}{(\sinh Py)^2}$, which is expressed as
\begin{align}
P \frac{\sinh ly}{(\sinh Py)^2} = -\frac{d}{dy} \bigg( \frac{\sinh (P+l)y}{\sinh Py} \bigg) + l \frac{\cosh (P+l)y }{\sinh Py}	\,	.	\label{rsinh}
\end{align}
In addition to \eqref{sinh}, we also have
\begin{align}
\frac{\cosh (P+l)y }{\sinh Py} &= \sum_{m=0}^{M_l} (e^{(l-2mP)y} - e^{-(l-2mP)y}) + \sum_{n=0}^{\infty} \varphi_{2P}^{(l-2M_l P)}(n) e^{-ny}	\,	,	\label{cosh}
\end{align}
for $M_l=-1,0,1, \ldots$ such that $2M_l P < l < 2(M_l+1)P$ where
\begin{align}
\varphi_{2P}^{(l)}(n) = 
\begin{cases}
1	&	\text{if } n \equiv \pm l \text{ mod } 2P	\\
0	&	\text{otherwise}
\end{cases}	\,	.
\end{align}
The values that $l$ in \eqref{rsinh} can take are
\begin{align}
P\bigg(\frac{1}{P_1}+\frac{1}{P_2}+\frac{1}{P_3}\bigg)	\,	,	\
P\bigg(\frac{1}{P_1}-\frac{1}{P_2}-\frac{1}{P_3}\bigg)		\,	,	\
P\bigg(-\frac{1}{P_1}+\frac{1}{P_2}-\frac{1}{P_3}\bigg)	\,	,	\
P\bigg(-\frac{1}{P_1}-\frac{1}{P_2}+\frac{1}{P_3}\bigg)	\,	,	\label{valueu}
\end{align}
where we denote them by $U_s$, $s=0,1,2,3$, respectively.
For all possible $(P_1,P_2,P_3)$ with the pairwise coprime condition, we see that $M_l=0$ or $-1$. 
Thus,
\begin{align}
P \frac{\sinh ly}{(\sinh Py)^2} = \sum_{n=0}^\infty \big(-n \, \psi_{2P}^{(l)}(n) + l \, \varphi_{2P}^{(l)}(n) \big) e^{-ny} =: \sum_{n=0}^\infty B_{2P}^{(l)}(n) e^{-ny}	\,	.	\label{expa}
\end{align}
where we introduce a notation $B_{2P}^{(l)}(n)$
\begin{align}
B_{2P}^{(l)}(n) := -n \, \psi_{2P}^{(l)}(n) + l \, \varphi_{2P}^{(l)}(n)	\,	.
\end{align}
We can also check directly that  
\begin{align}
(e^{Py} - e^{-Py}) \sum_{n=0}^\infty B_{2P}^{(l)}(n) e^{-ny} = 2P \sum_{n=0}^{\infty} \psi_{2P}^{(P-l)}(n) e^{-ny}
\end{align}
holds, which is consistent with \eqref{expa}.

From \eqref{su2int} and \eqref{expa}, we obtain the homological block $F(M_3,K;x,q)$
\begin{align}
F(M_3,K;x,q) = \sum_{n=0}^{\infty} (x^{n/2} - x^{-n/2}) \mathcal{B}_{2P}(n) q^{\frac{n^2}{4P}}	
\label{su2hb1}
\end{align}
where
\begin{align}
x:= e^{2\pi i u} = q^R	\,	, 	\quad	u:=\frac{R}{k}
\end{align}
and
\begin{align}
\mathcal{B}_{2P}(n) := \sum_{s=0}^{3} B_{2P}^{(U_s)}(n)	\,	.
\end{align}
In order to obtain the Wilson loop operator expectation value \eqref{su2full} from the homological block \eqref{su2hb1}, an overall factor $\frac{1}{8 i P} \sqrt{\frac{2}{k}} q^{-\frac{1}{4}(\theta_0-P)}$ should be multiplied to \eqref{su2hb1}.

This homological block can be understood as a contribution from the abelian flat connection to the partition function of the analytically continued Chern-Simons theory on Seifert knot complement, which is realized as M5$^{\prime}$ branes in the brane realization \eqref{mconf2} of the Chern-Simons theory.
The variable $x$ is understood as one of eigenvalues of the holonomy along the meridian of a knot complement.

\subsubsection*{Another way of calculation}

We have calculated the homological block by directly evaluating \eqref{su2int}, which contains a Gaussian factor.
The integral with such a Gaussian factor can be expressed as another integral expression, 
\begin{align}
\int_{\gamma} dy \, e^{-\frac{k}{2\pi i} p y^2 } e^{-ny} = \pi \Big( \frac{2 i}{k} \frac{1}{p} \Big)^{\frac{1}{2}} \oint_{|z|=1} \frac{dz}{2\pi i z}  \sum_{m \in \mathbb{Z}} q^{\frac{m^2}{4p}} z^m z^{n}	\,	.	\label{intrel1}
\end{align}
With \eqref{intrel1}, the Gaussian integral \eqref{su2int} can be expressed as
\begin{align}
\Big( \frac{P}{2 i k}\Big)^{\frac{1}{2}} \oint \frac{dz}{2\pi i z}  \, \sum_{m \in \mathbb{Z}} q^{\frac{m^2}{4P}} z^{-m} \frac{z^{PR}-z^{-PR}}{z^{P}-z^{-P}} \frac{\prod_{j=1}^{F} z^{P/P_j} - z^{-P/P_j}}{(z^{P} - z^{-P})^{F-2}}	\,	.	\label{su2int2}
\end{align}
When evaluating this integral, we take an expansion of the rational function as a series of $z$ as done in the previous calculation in terms of $e^{-y}$.
When $R$ is an integer, by using \eqref{sinh} and \eqref{char}, we obtain the same result as \eqref{intr2}.
We can also obtain the same result by using the expansion \eqref{expa} in \eqref{su2int2}.	\\

When $R$ is analytically continued, we can also calculate the homological block $F(M_3,K;x,q)$ in a similar way.
For $H=1$ and $F=3$, \eqref{su2int} can be expressed as
\begin{align}
\frac{P}{2\pi i} e^{\frac{\pi i}{2} k u^2 }\int_{\gamma} dy \, \Big( e^{-\frac{k}{2\pi i} P (y-\pi i u)^2} - e^{-\frac{k}{2\pi i} P (y+\pi i u)^2} \Big) \frac{\prod_{j=1}^{F} e^{Py/P_j} - e^{-Py/P_j}}{(e^{Py} - e^{-Py})^{2}}	\,	.	\label{su2int3}
\end{align}
By applying
\begin{align}
\int_{\gamma} dy \, e^{-\frac{k}{2\pi i} p (y \mp \pi i u)^2 } e^{-ny} = \pi \Big( \frac{2 i}{k} \frac{1}{p} \Big)^{\frac{1}{2}} \oint_{|z|=1} \frac{dz}{2\pi i z} \sum_{m \in \mathbb{Z}} q^{\frac{m^2}{4p}} z^m x^{\pm \frac{m}{2}} z^{n}	\label{intrel2}
\end{align}
to \eqref{su2int3}, we obtain \eqref{su2hb1}.\footnote{
One may also consider yet another way of calculation from \eqref{su2int} by taking
\begin{align}
e^{-Ry} = \frac{\theta(e^{-y};q) \theta(q^R;q)}{\theta(q^R e^{-y};q)\theta(1;q)}	\quad	\leadsto	\quad	\frac{\theta(z;q) \theta(x;q)}{\theta(x z;q)\theta(1;q)}	\label{rtheta}
\end{align}
where
\begin{align}
\theta(x;q)=(-q^{1/2}x;q)_\infty (-q^{1/2}x^{-1};q)_\infty	\,	.
\end{align}
However, \eqref{rtheta} is a series of $q^{1/2}$, not $q$, so it is not appropriate.
}	\\

As a remark, resurgent analysis is more apparent in the integral expression involving a Gaussian term.
Resurgent analysis is less apparent in the calculation from the contour integral involving theta-function-like term in the integrand, but it shows the structure more directly, which is useful for the case of $H \geq 2$.	\\

Calculations so far are for a loop supported on the fiber at a regular point on the orbifold.
For a loop on the fiber at the orbifold point of the weight $P_j$, we take $y \rightarrow y/P_j$ in the character $\chi_R(y)$ \cite{Blau-Thompson-Seifert}.	\\

We also note that all representations, \textit{i.e.} both odd and even $R$ gave non-zero results in the calculation above. 
However, in \cite{Gukov-Pei-Putrov-Vafa}, for lens space, only odd $R$'s provide non-zero homological blocks.
This is the case if we assume that only
\begin{align}
\int_{\gamma} dy \, e^{-\frac{k}{2\pi i} p y^2 } e^{-2ny} = \pi \Big( \frac{2 i}{k} \frac{1}{p} \Big)^{\frac{1}{2}} \oint_{|z|=1} \frac{dz}{2\pi i z}  \sum_{m \in \mathbb{Z}} q^{\frac{m^2}{4p}} z^m z^{2n}	\label{intrel2}
\end{align}
can be used for the calculation and \eqref{intrel1} is not allowed in the calculation.
But from the perspective of analytic continuation or resurgent analysis, there is no reason to exclude the case of even $R$.
We discuss this more in appendix \ref{sec:lens}.

\subsection{The case with more abelian flat connections}
\label{ssec:su2rshs}

As in the previous section, we consider both cases that $R$ is an integer and is analytically continued.
As will be discussed below, we see a difference between them.

\subsubsection*{Integer $R$}
From \eqref{su2int}, when $F=3$, we have
\begin{align}
\frac{1}{2\pi i} \sum_{t=0}^{H-1} e^{2 \pi i k \frac{P}{H}t^2}\int_{\gamma} dy \, e^{-\frac{k}{2\pi i} \frac{H}{P} (y+ 2\pi i \frac{P}{H}t)^2} \chi_R(y) \frac{\prod_{j=1}^{3} e^{y/P_j} - e^{-y/P_j}}{e^y - e^{-y}}	\,	.
\label{su2inth}
\end{align}
We first directly evaluate the integral and obtain
\begin{align}
\Big( \frac{1}{2 i k} \frac{H}{P} \Big)^{\frac{1}{2}} \sum_{t=0}^{H-1} \sum_{j=0}^{R-1} \sum_{n=0}^{\infty} \mathcal{A}_{2P}(n) q^{\frac{1}{4HP} (-2kPt +(R-1-2j)P-n)^2}
\end{align}
and when $(P_1,P_2,P_3)=(2,3,5)$ there is also 
\begin{align}
\sum_{t=0}^{H-1} \sum_{j=0}^{R-1} \Big( q^{\frac{1}{120H} (-60kt+30(R-1-2j)+1)^2} + q^{\frac{1}{120H} (-60kt+30(R-1-2j)-1)^2} \Big)	\,	.
\end{align}
Taking the limit $q \searrow e^{2\pi i /k}$ gives the Wilson loop expectation value, but the structure with the $S$-transform is not obvious in this calculation, so we may change the integral above to the contour integral.	\\

By using
\begin{align}
\int_{\gamma} dy \, e^{-\frac{k}{2\pi i} HP \big( y + 2\pi i \frac{t}{H} \big)^2 } e^{-ny} = \pi \Big( \frac{2 i}{k} \frac{1}{HP} \Big)^{\frac{1}{2}}  \oint_{|z|=1} \frac{dz}{2\pi i z} \sum_{m \in \mathbb{Z}} q^{\frac{m^2}{4HP}} z^{m} e^{-2 \pi i \frac{tm}{H}} z^{n}	\,	,	\label{intrel3}
\end{align}
\eqref{su2inth} can be expressed as
\begin{align}
\Big( \frac{1}{2 i k} \frac{P}{H} \Big)^{\frac{1}{2}} \sum_{t=0}^{H-1} e^{2\pi i k \frac{P}{H} t^2}\oint \frac{dz}{2\pi i z}  \, \sum_{m \in \mathbb{Z}} q^{\frac{m^2}{4HP}} z^{-m} e^{2\pi i \frac{tm}{H}} \frac{z^{PR}-z^{-PR}}{z^{P}-z^{-P}} \frac{\prod_{j=1}^{F} z^{P/P_j} - z^{-P/P_j}}{z^{P} - z^{-P}}	\,	.	\label{su2inth2}
\end{align}
In \eqref{su2inth2}, we may take
\begin{align}
\sum_{m \in \mathbb{Z}} = \sum_{b \in \mathbb{Z}/2H\mathbb{Z}} \sum_{l \in 2H\mathbb{Z} +b}	\,		\label{dsum}
\end{align}
then we have
\begin{align}
\begin{split}
&Z(k,R) \simeq  \Big( \frac{1}{2 i k} \frac{P}{H} \Big)^{\frac{1}{2}} \sum_{t=0}^{H-1} e^{2\pi i k \frac{P}{H} t^2} \sum_{b \in \mathbb{Z}/2H\mathbb{Z}} e^{2\pi i \frac{1}{H} tb} \\
&\hspace{40mm} \times \oint \frac{dz}{2\pi i z}  \, \sum_{l \in 2H\mathbb{Z}+b} q^{\frac{l^2}{4HP}} z^{-l} \sum_{j=0}^{R-1} z^{(R-1-2j)P} \sum_{n=0}^{\infty} \mathcal{A}_{2P}(n) z^{n} \ \Bigg|_{q \searrow e^{\frac{2\pi i }{k}}}	\,	.
\end{split}	\label{su2inth3}
\end{align}
Or we keep $z^{PR}-z^{-PR}$ and expand $\frac{\prod_{j=1}^{F} z^{P/P_j} - z^{-P/P_j}}{(z^{P} - z^{-P})^2}$, which leads to
\begin{align}
\begin{split}
&Z(k,R) \simeq \Big( \frac{1}{2 i k} \frac{P}{H} \Big)^{\frac{1}{2}} \sum_{t=0}^{H-1} e^{2\pi i k \frac{P}{H} t^2} \sum_{b \in \mathbb{Z}/2H\mathbb{Z}} e^{2\pi i \frac{1}{H} tb} \\
&\hspace{50mm}\times \oint \frac{dz}{2\pi i z}  \, \sum_{l \in 2H\mathbb{Z}+b} q^{\frac{l^2}{4HP}} z^{-l} (z^{PR}-z^{-PR}) \sum_{n=0}^{\infty} \mathcal{B}_{2P}(n) z^{n} \ \Bigg|_{q \searrow e^{\frac{2\pi i }{k}}}	\,	.
\end{split}	\label{su2inth31}
\end{align}
We could also take $\sum_{m \in \mathbb{Z}} = \sum_{b \in \mathbb{Z}/H\mathbb{Z}} \sum_{l \in H\mathbb{Z} +b}$, which is still fine to take $e^{2\pi i \frac{1}{H} tm}$ out of the integral.
But, \eqref{dsum} is a better choice, because depending on $P_j$'s the value of $Y_j$'s in \eqref{valuey} or $U_j$'s in \eqref{valueu}, \textit{i.e.} the values of $n$ that give non-zero $\mathcal{A}_{2P}(n)$ or $\mathcal{B}_{2P}(n)$, are either all odd or all even.

Also, \eqref{su2inth} can be expressed as
\begin{align}
\begin{split}
&\frac{P}{2\pi i} \sum_{t=0}^{H-1} e^{2\pi i k \frac{P}{H}t^2 + \frac{\pi i}{2k} \frac{P}{H} R^2 } \int_{\gamma} dy \, ( e^{-2 \pi i \frac{P}{H}t R} e^{-\frac{k}{2\pi i} HP (y+ 2\pi i \frac{t}{H} - \pi i \frac{R}{kH})} - e^{2 \pi i \frac{P}{H}t R} e^{-\frac{k}{2\pi i} HP (y+ 2\pi i \frac{t}{H} + \pi i \frac{R}{kH})})	\\
&\hspace{80mm}\times \frac{\prod_{j=1}^{3} e^{Py/P_j} - e^{-Py/P_j}}{(e^{Py} - e^{-Py})^2}		\,	,	\label{su2inth32}
\end{split}
\end{align}
and by using
\begin{align}
\int_{\gamma} dy \, e^{-\frac{k}{2\pi i} HP \big( y + 2\pi i \frac{t}{H} \pm \pi i \frac{R}{kH} \big)^2 } e^{-ny} = \pi \Big( \frac{2 i}{k} \frac{1}{HP} \Big)^{\frac{1}{2}}  \oint_{|z|=1} \frac{dz}{2\pi i z} \sum_{m \in \mathbb{Z}} q^{\frac{m^2}{4HP}} z^{-m} e^{2 \pi i \frac{tm}{H}} e^{\pm \pi i \frac{mR}{kH}} z^{n}		\,	,	\label{intrel4wr}
\end{align}
we obtain
\begin{align}
\begin{split}
&Z(k,R) \simeq -\Big( \frac{1}{2 i k} \frac{P}{H} \Big)^{\frac{1}{2}}  
\sum_{t=0}^{H-1} e^{2\pi i k \frac{P}{H}t^2} e^{\frac{\pi i}{2k} \frac{P}{H} R^2 }
\sum_{b \in \mathbb{Z}/2H\mathbb{Z}} e^{2 \pi i \frac{tb}{H}}
\oint \frac{dz}{2\pi i z} \sum_{l \in 2H\mathbb{Z}+b} q^{\frac{l^2}{4HP}} z^{-l}  	\\
&\hspace{60mm}\times (e^{2 \pi i \frac{P}{H}t R} e^{\pi i \frac{lR}{kH}} - e^{-2 \pi i \frac{P}{H}t R} e^{-\pi i \frac{lR}{kH}})
\sum_{n=0}^{\infty} \mathcal{B}_{2P}(n) z^{n} \Big|_{q \searrow e^{\frac{2\pi i }{k}}}		\,	.
\end{split}	\label{su2inth33}
\end{align}
Though \eqref{su2inth3}, \eqref{su2inth31}, and \eqref{su2inth33} look different, they are all the same.

From \eqref{su2inth3}, the homological block is given by
\begin{align}
\widehat{Z}_b = \frac{1}{|\mathcal{W}_b|} \oint \frac{dz}{2\pi i z}  \, \bigg( \sum_{l \in 2H\mathbb{Z}+b} q^{\frac{l^2}{4HP}} z^{-l} + \sum_{l \in 2H\mathbb{Z}-b} q^{\frac{l^2}{4HP}} z^{-l} \bigg) \sum_{j=0}^{R-1} z^{(R-1-2j)P} \sum_{n=0}^{\infty} \mathcal{A}_{2P}(n) z^{n}	\,	,	\label{su2hb2}
\end{align}
or from \eqref{su2inth31}
\begin{align}
\widehat{Z}_b = \frac{1}{|\mathcal{W}_b|} \oint \frac{dz}{2\pi i z}  \, \bigg( \sum_{l \in 2H\mathbb{Z}+b} q^{\frac{l^2}{4HP}} z^{-l} + \sum_{l \in 2H\mathbb{Z}-b} q^{\frac{l^2}{4HP}} z^{-l} \bigg) (z^{PR}-z^{-PR}) \sum_{n=0}^{\infty} \mathcal{B}_{2P}(n) z^{n}	\label{su2hb3}
\end{align}
and $e^{2\pi i \frac{1}{H} tb}$ in \eqref{su2inth3} and \eqref{su2inth31} contributes to the $S$-transform.
When $\chi_{R=1}(y)=1$, we see that \eqref{su2inth3} or \eqref{su2inth31} produces the result without a knot.
The label $b$ in \eqref{su2hb2} and in \eqref{su2hb3} are not the same label used for the case without a knot.
Instead, it is shifted due to the presence of a knot or the character, which we still denote as $b$.\footnote{We refer to section \ref{sec:hr} for the expression with the shifted label.}
\footnote{This can be expected from the M-theory configuration \eqref{mconf1} \cite{Gukov-Pei-Putrov-Vafa}.
The homological block counts BPS particles of $T[M_3]$ that is provided by M2 branes ending on M5 branes where the boundary condition is given by the torsion 1-cycle $(\tilde{b},-\tilde{b})$ with $[\tilde{b}] = b \in \text{Tor} H_1(M_3,\mathbb{Z})/\mathbb{Z}_2$.
The Wilson loop operator under consideration is also provided by M2 branes as in \eqref{mconf1}. 
It is expected to be supported on the knot $K$ that wraps a torsion 1-cycle as above, while the representation is given by the partition of the number of M2 branes into the number of M5 branes.
Therefore, it is expected that $b$ is affected by the presence of the Wilson loop operator.
}

Thus, we see that the structure of Chern-Simons partition function for Seifert knots $K$ with representation $R$ is given by
\begin{align}
Z_{SU(2)}(M_3, K; R,k) \simeq \sum_{a,b \in \text{Tor }H_1(M_3,\mathbb{Z})/\mathbb{Z}_2} e^{2\pi i k CS_a} \, S_{ab} \, \widehat{Z}_b(K;R,q) \Big|_{q \searrow e^{\frac{2\pi i}{k}}}	\,	.		\label{strm2}
\end{align}

For example, when $(P_1,P_2,P_3)=(2,3,7)$ with $H=5$ and $R=4$, we have
\begin{align}
\begin{split}
Z_{SU(2)}(M_3, K; 4,k) 
\simeq
\sum_{a,b=0}^{1} e^{2 \pi i k CS_a }S_{ab} \widehat{Z}_{b}
\end{split}
\end{align}
with $(CS_0,CS_1)=(0,\frac{1}{3})$, $S_{ab} =\frac{1}{\sqrt{3}} 
\begin{pmatrix}
1	&	1	\\
2	&	-1
\end{pmatrix}
$,
and
\begin{align}
\widehat{Z}_0 &= \widetilde{\Psi }_{210}^{(39)}+\widetilde{\Psi }_{210}^{(81)}+\widetilde{\Psi }_{210}^{(129)}+\widetilde{\Psi }_{210}^{(171)}	\,	,	\\
\widehat{Z}_1 &= -\widetilde{\Psi }_{210}^{(11)}+\widetilde{\Psi }_{210}^{(31)}-\widetilde{\Psi }_{210}^{(59)}-\widetilde{\Psi }_{210}^{(101)}-\widetilde{\Psi}_{210}^{(109)}-\widetilde{\Psi }_{210}^{(151)}+\widetilde{\Psi }_{210}^{(179)}-\widetilde{\Psi }_{210}^{(199)}	\,	,
\end{align}
where $\widetilde{\Psi}_P^{(l)}(q) = \sum_{n=0}^{\infty}\psi_{2P}^{(l)}(n) q^{\frac{n^2}{4P}}$ is a false theta function \cite{Lawrence-Zagier}.

For a knot on the fiber at the orbifold point of weight $P_j$'s, homological blocks are given by, respectively,	\\
$P_1=2$
\begin{align}
\widehat{Z}_0 &= 0	\,	,\hspace{109mm}	\\
\widehat{Z}_1 &= 0	\,	,\hspace{109mm}
\end{align}
$P_2=5$
\begin{align}
\widehat{Z}_0 &= -\widetilde{\Psi }_{210}^{(39)}-\widetilde{\Psi }_{210}^{(81)}-\widetilde{\Psi }_{210}^{(171)}+\widetilde{\Psi }_{210}^{(291)}	\,	,	\\
\widehat{Z}_1 &= \widetilde{\Psi }_{210}^{(59)}+\widetilde{\Psi }_{210}^{(101)}+\widetilde{\Psi }_{210}^{(109)}+\widetilde{\Psi }_{210}^{(151)}+\widetilde{\Psi
   }_{210}^{(199)}+\widetilde{\Psi }_{210}^{(241)}+\widetilde{\Psi }_{210}^{(389)}-\widetilde{\Psi }_{210}^{(409)}	\,	,
\end{align}
$P_3=7$
\begin{align}
\widehat{Z}_0 &= -\widetilde{\Psi }_{210}^{(9)}-\widetilde{\Psi }_{210}^{(51)}-\widetilde{\Psi }_{210}^{(159)}+\widetilde{\Psi }_{210}^{(219)}	\,	,	\\
\widehat{Z}_1 &= -\widetilde{\Psi }_{210}^{(61)}+\widetilde{\Psi }_{210}^{(79)}+\widetilde{\Psi }_{210}^{(89)}+\widetilde{\Psi }_{210}^{(121)}+\widetilde{\Psi}_{210}^{(131)}+\widetilde{\Psi }_{210}^{(229)}+\widetilde{\Psi }_{210}^{(271)}+\widetilde{\Psi }_{210}^{(401)}	\,	.
\end{align}
\vspace{0mm}

\subsubsection*{Analytically continued $R$}

The case with an integer $R$ was similar to the case of 3-manifolds without a knot in a sense that the Wilson loop expectation value is decomposed into homological blocks as in \eqref{strm2}.
However for the case of analytically continued $R$, we see that the decomposition as in \eqref{strm2} doesn't arise except the case of a single abelian flat connection $H=1$ where the $S$-transform is trivial.	\\

If directly evaluating the integral with a Gaussian factor, from \eqref{su2int} we have
\begin{align}
\begin{split}
&\frac{P}{2\pi i} \sum_{t=0}^{H-1} e^{2\pi i k \frac{P}{H}t^2 + \frac{\pi i}{2}k \frac{P}{H} u^2 }  \int_{\gamma} dy \, ( e^{-2 \pi i k \frac{P}{H}t u} e^{-\frac{k}{2\pi i} HP (y+ 2\pi i \frac{t}{H} - \pi i \frac{u}{H})^2} + e^{2 \pi i k \frac{P}{H}t u} e^{-\frac{k}{2\pi i} HP (y+ 2\pi i \frac{t}{H} + \pi i \frac{u}{H})^2})	\\
&\hspace{52mm}\times \frac{\prod_{j=1}^{3} e^{Py/P_j} - e^{-Py/P_j}}{(e^{Py} - e^{-Py})^2}	\,	,	\label{su2inthr10}
\end{split}
\end{align}
and this gives
\begin{align}
Z(k,R) \simeq -\sum_{t=0}^{H-1} \sum_{n=0}^{\infty} \mathcal{B}_{2P}(n) \big( e^{2\pi i k \frac{P}{H} (t+ \frac{u}{2})^2} x^{\frac{n}{2H}} - e^{2\pi i k \frac{P}{H} (t - \frac{u}{2})^2} x^{-\frac{n}{2H}} \big) e^{2\pi i \frac{nt}{H}} q^{\frac{n^2}{4HP}} \Big|_{q \searrow e^{\frac{2\pi i }{k}}, \, x\rightarrow e^{2\pi i \frac{R}{k}}}	\label{hbfint}
\end{align}
up to an overall factor.

Or by using
\begin{align}
\int_{\gamma} dy \, e^{-\frac{k}{2\pi i} HP \big( y + 2\pi i \frac{t}{H} \pm \pi i \frac{u}{H} \big)^2 } e^{-ny} = \pi \Big( \frac{2 i}{k} \frac{1}{HP} \Big)^{\frac{1}{2}}  \oint_{|z|=1} \frac{dz}{2\pi i z} \sum_{m \in \mathbb{Z}} q^{\frac{m^2}{4HP}} z^{-m} e^{2 \pi i \frac{tm}{H}} x^{\pm \frac{m}{2H}} z^{n}		\,	,	\label{intrel4}
\end{align}
we obtain
\begin{align}
\begin{split}
&Z(k,R) \simeq -\Big( \frac{1}{2 i k} \frac{P}{H} \Big)^{\frac{1}{2}}  
\sum_{t=0}^{H-1} e^{2\pi i k \frac{P}{H}t^2 + \frac{\pi i}{2}k \frac{P}{H} u^2 }
\sum_{b \in \mathbb{Z}/2H\mathbb{Z}} e^{2 \pi i \frac{tb}{H}}
\oint_{|z|=1} \frac{dz}{2\pi i z} \sum_{l \in 2H\mathbb{Z}+b} q^{\frac{l^2}{4HP}} z^{-l}  	\\
&\hspace{60mm}\times (e^{2 \pi i k \frac{P}{H}t u} x^{\frac{l}{2H}} - e^{-2 \pi i k \frac{P}{H}t u} x^{-\frac{l}{2H}})
\sum_{n=0}^{\infty} \mathcal{B}_{2P}(n) z^{n} \Big|_{q \searrow e^{\frac{2\pi i }{k}}, \, x\rightarrow e^{2\pi i \frac{R}{k}}}		\,	.
\end{split}	\label{su2inthr1}
\end{align}
This agrees with \eqref{su2inth33} when $R$ is taken to be an integer.	\\

The sum of the contributions from $t$ and $H-t$ is
\begin{align}
\begin{split}
&-\Big( \frac{1}{2 i k} \frac{P}{H} \Big)^{\frac{1}{2}}  
e^{2\pi i k \frac{P}{H}t^2 + \frac{\pi i}{2}k \frac{P}{H} u^2 }
\sum_{b \in \mathbb{Z}/2H\mathbb{Z}}
\oint_{|z|=1} \frac{dz}{2\pi i z} \sum_{l \in 2H\mathbb{Z}+b} q^{\frac{l^2}{4HP}} z^{-l} \sum_{n=0}^{\infty} \mathcal{B}_{2P}(n) z^{n}	\\
&\hspace{55mm}\times 
\Big( e^{2 \pi i \frac{tb}{H}} (e^{2 \pi i \frac{P}{H}t R} x^{\frac{l}{2H}} - e^{-2 \pi i \frac{P}{H}t R} x^{-\frac{l}{2H}})	\\
&\hspace{65mm}+e^{-2 \pi i \frac{tb}{H}} (e^{2 \pi i \frac{P}{H}(H-t) R} x^{\frac{l}{2H}} - e^{-2 \pi i \frac{P}{H}(H-t) R} x^{-\frac{l}{2H}})
\Big)
 \Big|_{q \searrow e^{\frac{2\pi i }{k}}, \, x\rightarrow e^{2\pi i \frac{R}{k}}}
\end{split}	\label{su2inthr2}
\end{align}
where we have already taken $k$ to be an integer at $e^{2\pi i k \frac{P}{H}t^2}$. 
We see that $e^{\pm 2\pi i \frac{P}{H} tR} = e^{\pm 2 \pi i k \frac{P}{H} t u}$ prevents to have a common factor $e^{2\pi i \frac{tb}{H}}+e^{-2\pi i \frac{tb}{H}}$ in \eqref{su2inthr2}.
This coupling between $u$ and $t$ is always present when $H \geq 2$.
So, for the Seifert rational homology sphere, when $R$ is analytically continued it is not possible to express the Chern-Simons partition function with a Wilson loop operator on the Seifert knot in terms of the $(x,q)$-series with integer coefficients while having the structure with the $S$-transform.
We give a few more remarks in Appendix \ref{sec:rmk}.

\section{Higher rank case}
\label{sec:hr}

For a Seifert knot, the contributions from the abelian flat connections to the partition function has been obtained in \cite{Blau-Thompson-Seifert} for simply-laced group $G$.\footnote{The contribution from the trivial flat connection has been obtained in \cite{Beasley-Wilson} for the compact, connected, simply-connected, and simple gauge group $G$.}
As done in the case of Seifert manifolds without a knot \cite{Chung-Seifert} by using the expressions of the Chern-Simons partition function in \cite{Marino2004}, we can obtain the homological block for the Seifert knot from the expression in \cite{Blau-Thompson-Seifert}.
Here, we consider the case of $G=SU(N)$ and other types of simply-laced group can be done similarly.	\\

When $G=SU(N)$, the contribution from the abelian flat connection is given by
\begin{align}
\begin{split}
\sum_{{\bf t} \in \Lambda_{\text{rt}} / H \Lambda_{\text{rt}}}	\int_{\Gamma_{\mathbf{t}}^{N-1}} d\beta_1 \ldots d\beta_{N-1} 
& \, e^{-\frac{k}{2 \pi i} \frac{H}{P} (\sum_{i=1}^{N-1} \beta_i^2 + \sum_{i<j}^{N-1} \beta_i \beta_j) - k (2 \sum_{i=1}^{N-1} t_i \beta_i + \sum_{i \neq j}^{N-1} t_i \beta_j )}	\\
& \qquad	\qquad	\times \text{ch}_\mathcal{R}(e^\beta) \ \frac{\prod_{f=1}^{F} \prod_{i<j}^{N} 2 \sinh \frac{1}{2 P_f} (\beta_i - \beta_j) }{\prod_{i<j}^{N} \big(2 \sinh \frac{1}{2} (\beta_i - \beta_j) \big)^{F-2}}
\end{split}	\label{sunint1}
\end{align}
up to the overall factor.\footnote{\label{ov-factor} When $G=SU(N)$ and there is no knot, the overall factor of the integral is
\begin{align}
\frac{(-1)^{(N-1)N/2}}{(2 \pi i)^{N-1} N!} \frac{1}{N} \frac{\text{sign}(P)^{\frac{N(N-1)}{2}} }{|P|^{\frac{N-1}{2}}} e^{\frac{\pi i (N^2-1)}{4} \text{sign}(H/P) - \frac{\pi i}{12 k} (N^2-1) N \phi_F }
\end{align}
where $\phi_F	= 3 \, \text{sign}\left( \frac{H}{P} \right) + \sum_{j=1}^{F} \left( 12 s(Q_j, P_j) - \frac{Q_j}{P_j} \right)$.
For the case with a knot, the overall factor was not determined in \cite{Blau-Thompson-Seifert}.
In this section we work on the expressions up to an overall factor. 
} 
Here, $\Gamma_{\mathbf{t}}^{N-1}$ denotes the contour for each $\beta_j$, $j=1, \ldots, N-1$, that passes a stationary phase point as the contour $C_t$ in \eqref{su2full}.
For a knot on the fiber at the orbifold point of the weight $P_j$, the character $\text{ch}_R(e^\beta)$ is replaced by $\text{ch}_R(e^{\beta/P_j})$.

\subsubsection*{Integer $\mathcal{R}$}

The representation $\mathcal{R}$ is labelled by Young tableaux, which is a set of integers, and the character is a polynomial.
As done previously, we consider an expansion of $\prod_{f=1}^{F} \prod_{i<j}^{N} 2 \sinh \frac{1}{2 P_f} (\beta_i - \beta_j) /\prod_{i<j}^{N} \big(2 \sinh \frac{1}{2} (\beta_i - \beta_j) \big)^{F-2}$.
For concreteness, we consider the case of $F=3$ and other cases can be done similarly.	\\

For $i$ and $j$ with $i<j$ and $j \neq N$, we have
\begin{align}
\frac{\prod_{f=1}^{3} \Big( e^{ \frac{1}{P_f}\frac{1}{2} (\beta_i - \beta_j)} - e^{-\frac{1}{P_f}\frac{1}{2} (\beta_i - \beta_j)} \Big)}{e^{\frac{1}{2} (\beta_i - \beta_j)} - e^{-\frac{1}{2} (\beta_i - \beta_j)}}
= \sum_{n_{i,j}=0}^{\infty} \mathcal{A}_{2P}(n_{i,j}) \, e^{\frac{1}{2P} n_{i,j} (\beta_{i} - \beta_{j})}	\,	,	\label{sunexp1}
\end{align}
and for $j=N$,
\begin{align}
\frac{\prod_{f=1}^{3} \Big( e^{ \frac{1}{P_f}\frac{1}{2} (\beta_i - \beta_N)} - e^{-\frac{1}{P_f}\frac{1}{2} (\beta_i - \beta_N)} \Big)}{ e^{\frac{1}{2} (\beta_i - \beta_N)} - e^{-\frac{1}{2} (\beta_i - \beta_N)}}
= \sum_{n_{i,N}=0}^{\infty} \mathcal{A}_{2P}(n_{i,N}) \, e^{-\frac{1}{2P} n_{i,N} (\beta_{i} + \sum_{l=1}^{N-1} \beta_{l})}	\,		\label{sunexp2}
\end{align}
where we chose $0 < \text{Re} \, \beta_1 < \text{Re} \, \beta_2 < \cdots < \text{Re} \, \beta_{N-1}$ and $P>0$ for convergence.
We analytically continue the level $k$ and take a contour $\gamma_j$ for each $\beta_j$, $j=1,2, \ldots, N-1$ as the contour $\gamma$ for $y$ in \eqref{su2int}. 
With the expansions \eqref{sunexp1} and \eqref{sunexp2}, the integral \eqref{sunint1} becomes
\begin{align}
\begin{split}
\sum_{{\bf t} \in \Lambda_{\text{rt}} / H \Lambda_{\text{rt}}}
\int_{\gamma^{N-1}} \, \prod_{j=1}^{N-1} d\beta _j \,
&e^{-\frac{K}{4 \pi i} \frac{H}{P} \sum _{j=1}^{N} \beta _j^2 - K \sum _{i=1}^{N} t_j \beta _j } \, \text{ch}_R(e^\beta) \sum_{\substack{n_{i,j}=0 \\ 1 \leq i < j \leq N}}^{\infty} \Big( \prod_{1\leq i<j \leq N} \mathcal{A}_{2P}(n_{i,j}) \Big)	\,
e^{\frac{1}{2P} \sum _{j=1}^{N} c_j \beta _j } 
\end{split}	\label{sunint2}
\end{align}
where $\beta_N = -\sum_{j=1}^{N-1} \beta_j$, $t_N = -\sum_{j=1}^{N-1} t_j$, and
\begin{align}
c_i =  -\sum _{j=1}^{i-1} n_{j,i} + \sum _{j=i+1}^{N-1} n_{i,j} -n_{i,N} 	\,	,	\quad	
c_N =  \sum _{j=1}^{N-1} n_{j,N}	\,	.
\label{candn}
\end{align}
They satisfy $\sum_{j=1}^N c_j = 0$.
We also see that $\vec{c} = (c_1, \ldots, c_N) = \sum_{1\leq i < j \leq N-1} n_{i,j} e_{i,j} - \sum_{i=1} n_{i,N} e_{i,N}$ where $e_{i,j}$ is a $N$-component vector with $+1$ and $-1$ at the $i$-th and $j$-th component, respectively, and zero for the rest of components.
Thus, $\vec{c}$ is an element in the root lattice $\Lambda_{\text{rt}}$.\footnote{
Or the coroot lattice $\Lambda_{\text{cort}}$ as for $G=SU(N)$, $\Lambda_{\text{rt}}=\Lambda_{\text{cort}}$.
}

We can directly evaluate \eqref{sunint2}, or we can also use 
\begin{align}
\begin{split}
&\int_{\gamma^{N-1}} \prod_{j=1}^{N-1} d\beta_j \, e^{-\frac{k}{\pi i} HP \sum_{j=1}^N \big( \beta_j + \pi i \frac{t_j}{H} \big)^2 } \Big( \prod_{j=1}^{N} e^{d_j \beta_j} \Big)	\\
&\hspace{5mm} =  \frac{\pi^{N-1}}{\sqrt{N}} (i HP k)^{-\frac{N-1}{2}}
\oint_{|z|=1} \prod_{j=1}^{N-1} \frac{dz_j}{2\pi i z_j} \sum_{\vec{m} \in \mathbb{Z}^{N}} q^{\frac{1}{8NHP} \big( N \sum_{j=1}^{N} m_j^2 -(\sum_{j=1}^{N} m_j)^2 \big) } \prod_{j=1}^{N} z_j^{-m_j} 	\\
&\hspace{75mm} \times e^{\frac{\pi i}{H} \sum_{j=1}^{N} t_j m_j} \Big( \prod_{j=1}^{N} z_j^{-d_j} \Big)	\,	,	\label{intrelunr}
\end{split}
\end{align}
where $z_N = (\prod_{j=1}^{N-1}z_j)^{-1}$ and $\sum_{j=1}^N m_j = -\sum_{j=1}^N d_j$. 
For the trivial representation, \textit{i.e.} without a knot, \eqref{sunint2} can be expressed as
\begin{align}
\begin{split}
\hspace{-5mm} \sum_{{\bf t} \in \Lambda_{\text{rt}} / H \Lambda_{\text{rt}}} e^{\pi i k \frac{P}{H} \sum_{j=1}t_j^2}
\oint \, \prod_{j=1}^{N-1} \frac{dz_j}{2\pi i z_j} \,
&\sum_{\vec{m} \in \mathbb{Z}^N}q^{\frac{1}{8NHP} \big( N \sum_{j=1}^{N} m_j^2 - (\sum_{j=1}^{N} m_j)^2 \big) } \prod_{j=1}^{N} z_j^{m_j} e^{-\frac{\pi i}{H} \sum_{j=1}^{N}t_j m_j}	\\
&\hspace{-10mm}\times \sum_{\substack{n_{i,j}=0 \\ 1 \leq i < j \leq N}}^{\infty} \prod_{1\leq i<j \leq N} \mathcal{A}_{2P}(n_{i,j}) \prod _{j=1}^{N} z_j^{-c_j}
\end{split}	\label{sunint3}
\end{align}
where we have taken $\vec{m} \rightarrow -\vec{m}$ in the sum.
Since $\vec{m}$ takes values of $\vec{c}$ in \eqref{sunint3}, $\vec{m}$ is also in $\Lambda_{\text{rt}}$.
As $\vec{t}$ and $\vec{m}$ are in $\Lambda_{\text{rt}}$, we have $\vec{t} \cdot \vec{m} \in \mathbb{Z}$.
Therefore, in general, we may take
\begin{align}
\sum_{\vec{m} \in \mathbb{Z}^N} \, \rightarrow \sum_{\vec{b} \in \Lambda_{\text{rt}}/2H \Lambda_{\text{rt}}} \sum_{\vec{l} \in 2H \Lambda_{\text{rt}} + \vec{b}}	\label{sumdecomp}
\end{align}
and further express \eqref{sunint3} as 
\begin{align}
\begin{split}
&\sum_{{\bf t} \in \Lambda_{\text{rt}} / H \Lambda_{\text{rt}}} e^{\pi i k \frac{P}{H} \sum_{j=1}t_j^2} \sum_{\vec{b} \in \Lambda_{\text{rt}}/2H \Lambda_{\text{rt}}} e^{-\frac{\pi i}{H} \sum_{j=1}^{N} t_j b_j}	\\
&\hspace{30mm}\times \oint \, \prod_{j=1}^{N-1} \frac{dz_j}{2\pi i z_j} \, \sum_{\vec{l} \in 2H\Lambda_{\text{rt}}+ \vec{b}} q^{\frac{1}{8 N HP} \big( N \sum_{j=1}^{N} l_j^2 - (\sum_{j=1}^{N} l_j)^2 \big)} \prod_{j=1}^{N} z_j^{l_j} 	\\
&\hspace{50mm}\times \sum_{\substack{n_{i,j}=0 \\ 1 \leq i < j \leq N}}^{\infty} \prod_{1\leq i<j \leq N} \mathcal{A}_{2P}(n_{i,j}) \prod _{j=1}^{N} z_j^{-c_j}	
\end{split}	\label{sunint4}
\end{align}
where $e^{-\pi i \sum_{j=1}^N \frac{t_j b_j}{H}}$ contributes to the $S$-transform.

In \eqref{sunint4}, $\vec{b}$'s are supposed to label the homological blocks and isomorphic to $\mathbf{t}$, but $\vec{b} \in \Lambda_{\text{rt}}/2H \Lambda_{\text{rt}}$ while $\mathbf{t} \in \Lambda_{\text{rt}} / H \Lambda_{\text{rt}}$.
However, we see that they are indeed isomorphic.
From the conditions that $P_j$ and $Q_j$, $j=1, \ldots, F$, are coprime and $P_j$'s are pairwise coprime, $H$ and $P$ are coprime.
Meanwhile, $\mathcal{A}_{2P}(n_{i,j})$ take non-zero values when $n_{i,j}$'s take values in \eqref{valuey}.
Therefore, when $P$ is odd such $n_{i,j}$'s are all even, and when $P$ is even such $n_{i,j}$'s are all odd.

In \eqref{sunint3}, terms with $\vec{l}=\vec{c}$ survive where each $c_j$ is given by a linear combination of $N-1$ $n_{i,j}$'s from \eqref{candn}.
So when $n_{i,j}$'s are all even, regardless of $N$, all components of $\vec{c}$ are even, so are $\vec{l}$.
This is also the case when $n_{i,j}$'s are all odd and $N$ is odd.
Therefore, in this case, all components of $\vec{b}$ should also be even.
Such $\vec{b} \in \Lambda_{\text{rt}}/2H \Lambda_{\text{rt}}$ is obtained from multiplying $\mathbf{t} \in \Lambda_{\text{rt}} / H \Lambda_{\text{rt}}$ by 2.

If $P$ is even and $H$ is odd, the components of $n_{i,j}$'s are all odd.
When $N$ is even, all components of $\vec{c}$ are odd, so are $\vec{l}$.
Therefore, all components of $\vec{b}$ should also be odd.
We see that such $\vec{b} \in \Lambda_{\text{rt}}/2H \Lambda_{\text{rt}}$ can be obtained by $2 \mathbf{t} + H \sum_{j=1}^{N/2} \alpha_{2j-1,2j}$ where $\alpha_{i,j}$ denotes the root of $G=SU(N)$ where the $i$-th component is 1 and the $j$-th component is $-1$ in the orthonormal basis.

From the discussion above, the homological block $\widehat{Z}_{W_{\vec{b}}}(q)$ is given by
\begin{align}
\begin{split}
\widehat{Z}_{W_{\vec{b}}}(q) &= \frac{1}{|W_{\vec{b}}|} \sum_{\vec{b} \in W_{\vec{b}}} \oint \, \prod_{j=1}^{N-1} \frac{dz_j}{2\pi i z_j} \, \sum_{\vec{l} \in 2H\Lambda_{\text{rt}}+ \vec{b}}q^{\frac{1}{8NHP} \big( N \sum_{j=1}^{N} l_j^2 - (\sum_{j=1}^{N} l_j)^2 \big) } \prod_{j=1}^{N} z_j^{l_j}	\\
&\hspace{50mm} \times \sum_{\substack{n_{i,j}=0 \\ 1 \leq i < j \leq N}}^{\infty} \prod_{1\leq i<j \leq N} \mathcal{A}_{2P}(n_{i,j}) \prod _{j=1}^{N} z_j^{-c_j}	\,	.	\label{hbsft2}
\end{split}
\end{align}
where $W_{\vec{b}}$ denotes the Weyl orbit of $\vec{b}$ that satisfies the conditions discussed above.	\\

When considering a knot with a representation $\mathcal{R}$, by using \eqref{intrelunr}, \eqref{sunint2} can be expressed as
\begin{align}
\begin{split}
\hspace{-5mm} \sum_{{\bf t} \in \Lambda_{\text{rt}} / H \Lambda_{\text{rt}}} e^{\pi i k \frac{P}{H} \sum_{j=1}t_j^2}
\oint \, \prod_{j=1}^{N-1} \frac{dz_j}{2\pi i z_j} \,
&\sum_{\vec{m} \in \mathbb{Z}^N}q^{\frac{1}{8NHP} \big( N \sum_{j=1}^{N} m_j^2 - (\sum_{j=1}^{N} m_j)^2 \big) } \, \text{ch}_{\mathcal{R}}(z^{-2P}) \, \prod_{j=1}^{N} z_j^{m_j} e^{-\frac{\pi i}{H} \sum_{j=1}^{N}t_j m_j}	\\
&\hspace{0mm}\times \sum_{\substack{n_{i,j}=0 \\ 1 \leq i < j \leq N}}^{\infty} \prod_{1\leq i<j \leq N} \mathcal{A}_{2P}(n_{i,j}) \prod _{j=1}^{N} z_j^{-c_j}	\,	.
\end{split}	\label{sunint5}
\end{align}
In this case, due to the presence of the character $\vec{b}$ cannot take the value in the root space in general, so \eqref{sumdecomp} is not directly applied to \eqref{sunint5}.
Instead, given a knot with a representation $\mathcal{R}$, we consider monomials in $\text{ch}_\mathcal{R}(z^{-2P}) \prod _{j=1}^{N} z_j^{-c_j }$, which can be expressed as $h_{\vec{f}}\prod_{j=1}^{N} z_j^{-d_j}$ with $d_j = c_j+2P f_j$, $j=1, \ldots, N$ where a monomial in $\text{ch}_{\mathcal{R}}(z^{-2P})$ is denoted by $\prod_{j=1}^{N} z^{-2P f_j}$ and $h_{\vec{f}}$ is an integer.
Therefore, \eqref{sunint2} is expressed as 
\begin{align}
\begin{split}
&\hspace{-5mm} \sum_{{\bf t} \in \Lambda_{\text{rt}} / H \Lambda_{\text{rt}}} e^{\pi i k \frac{P}{H} \sum_{j=1}t_j^2}
\sum_{\vec{f}} h_{\vec{f}} \sum_{\vec{m} \in \mathbb{Z}} \oint \, \prod_{j=1}^{N-1} \frac{dz_j}{2\pi i z_j} \,
q^{\frac{1}{8NHP} \big( N \sum_{j=1}^{N} m_j^2 - (\sum_{j=1}^{N} m_j)^2 \big) } \prod_{j=1}^{N} z_j^{m_j} e^{-\frac{\pi i}{H} \sum_{j=1}^{N} t_j m_j}	\\
&\hspace{65mm}\times \sum_{\substack{n_{i,j}=0 \\ 1 \leq i < j \leq N}}^{\infty} \prod_{1\leq i<j \leq N} \mathcal{A}_{2P}(n_{i,j}) \prod _{j=1}^{N} z_j^{-d_j }
\end{split}	
\end{align}
where $\sum_{\vec{f}}$ denotes the sum over $\vec{f}=(f_1, \ldots, f_N)$ of the monomials.
This is also expressed as
\begin{align}
\begin{split}
&\hspace{0mm} \sum_{{\bf t} \in \Lambda_{\text{rt}} / H \Lambda_{\text{rt}}} e^{\pi i k \frac{P}{H} \sum_{j=1}t_j^2} \, \sum_{\vec{f}} h_{\vec{f}}  \sum_{\vec{b} \in \Lambda_{\text{rt}}/2H \Lambda_{\text{rt}}} e^{-\frac{\pi i}{H} \sum_{j=1}^{N} t_j (b_j + 2P f_j)}	\\
&\times \oint \, \prod_{j=1}^{N-1} \frac{dz_j}{2\pi i z_j} \, \sum_{\vec{l} \in 2H\Lambda_{\text{rt}}+ \vec{b}}q^{\frac{1}{8NHP} \big( N \sum_{j=1}^{N} (l_j+2Pf_j)^2 - (\sum_{j=1}^{N} 2P f_j)^2 \big) } \prod_{j=1}^{N} z_j^{l_j} \sum_{\substack{n_{i,j}=0 \\ 1 \leq i < j \leq N}}^{\infty} \prod_{1\leq i<j \leq N} \mathcal{A}_{2P}(n_{i,j}) \prod _{j=1}^{N} z_j^{-c_j}	\,	.
\end{split}	\label{sunint6}
\end{align}
Comparing \eqref{sunint6} with \eqref{sunint4} for the case without a knot, we see that $\vec{b}+2P\vec{f}$ serves an appropriate label for the homological block, which we denote as $[\vec{b}+2P\vec{f}]$.
Certainly, there are more numbers of $\vec{b}+2P\vec{f}$ than the number of abelian flat connections, but we see from examples that sets of $\vec{b}+2P\vec{f}$ appropriately label the homological blocks.
Thus, in the presence of knot, the homological block can be expressed as
\begin{align}
\begin{split}
&\widehat{Z}_{[\vec{b}+2P\vec{f}]}(K;q, R) = \sideset{}{'}\sum_{\vec{f}} h_{\vec{f}} \oint \, \prod_{j=1}^{N-1} \frac{dz_j}{2\pi i z_j} \, \sideset{}{'}\sum_{\vec{b} \in \Lambda_{\text{rt}}/2H \Lambda_{\text{rt}}}   \sum_{\vec{l} \in 2H\Lambda_{\text{rt}}+ \vec{b}} q^{\frac{1}{8NHP} \big( N \sum_{j=1}^{N} (l_j+2Pf_j)^2 - (\sum_{j=1}^{N} 2P f_j)^2 \big) } \prod_{j=1}^{N} z_j^{l_j} 	\\
&\hspace{40mm}\times \sum_{\substack{n_{i,j}=0 \\ 1 \leq i < j \leq N}}^{\infty} \prod_{1\leq i<j \leq N} \mathcal{A}_{2P}(n_{i,j}) \prod _{j=1}^{N} z_j^{-c_j}	\,	.
\end{split}	\label{hbsfk}
\end{align}
Here, the primed sums are the sum over $\vec{f}$ and $\vec{b}$ such that they are in the same class $[\vec{b}+2P\vec{f}]$ and this can be seen by checking the exponents of $q$.\footnote{Therefore the homological blocks for Seifert knots with the representation $\mathcal{R}$ calculated here are the version that is obtained when modding out flat connections by the action of the center and the complex conjugate \cite{Chung-Seifert}.}

\subsubsection*{Example}

When $G=SU(3)$, $H=2$, $(P_1,P_2,P_3)=(3,5,7)$, there are 5 Weyl orbits in $\Lambda_\text{rt}/4\Lambda_\text{rt}$, but only three Weyl orbits contribute
\begin{alignat}{3}
&W_{t}			&&\hspace{10mm}		\Lambda_\text{rt}/4\Lambda_\text{rt}	\nonumber	\\
\vspace{1mm}
&W_0 			&&\hspace{10mm}		\{0\}		\nonumber	\\
&W_1 			&&\hspace{10mm}		\{2\alpha _{12}, 2\alpha _{23}, 2(\alpha _{12}+\alpha _{23})\}	\nonumber	\\
&W_2 			&&\hspace{10mm}		\{4\alpha _{12}, 4\alpha _{23}, 4(\alpha _{12}+\alpha _{23})\}	\nonumber
\end{alignat}
where $\alpha_{12}=(1,-1,0)$, $\alpha_{13}=(1,0,-1)$, and $\alpha_{23}=(0,1,-1)$.
The Chern-Simons partition function with a knot is given by
\begin{align}
\sum_{a,b=0}^1 e^{2\pi i k CS_a} S_{ab} \widehat{Z}_b(M_3,K;R,q) \Big|_{q \searrow e^{\frac{2\pi i}{k}}}
\end{align}
where $(CS_0,CS_1)=(0,\frac{1}{2})$,
\begin{align}
S_{ab} = \frac{1}{2} \begin{pmatrix} 1 & 1 \\ 3 & -1\end{pmatrix}	\,	,
\end{align}
and, for example, for a totally symmetric representation $\mathcal{S}^2$, 
\begin{align}
\widehat{Z}_0(K;\mathcal{S}^2;q) &= q^{43+\frac{53}{105}} (-3-3 q^2-3 q^8+3 q^{22}-6 q^{32}+6 q^{34}+9 q^{46}-3 q^{48}+	\ldots )	\\	
\widehat{Z}_1(K;\mathcal{S}^2;q) &= q^{40+\frac{1}{210}} (3-3 q^{11}+3 q^{12}+3 q^{15}-3 q^{16}+6 q^{18}+6 q^{20}+6 q^{24}+	\ldots )	\,	.
\end{align}

When a knot is supported on the fiber at the orbifold point of the weight $P_j$, $j=1,2,3$, we take $\text{ch}_{\mathcal{R}}(e^{y/P_j})$ and obtain homological blocks, respectively,\\
$P_1=3$
\begin{align}
&\widehat{Z}_0 = q^{1+\frac{299}{315}} (2-2 q^4-2 q^6-2 q^8+q^{12}-q^{14}+q^{20}+3 q^{24}+q^{32}-q^{38}-q^{40} + \ldots )	\,	,	\\	
&\widehat{Z}_1 = -q^{283/630}(2-2 q^2-2 q^4+2 q^5+q^7-q^8-q^9-2 q^{11}-2 q^{14}+q^{15} + \ldots )		\,	,		
\end{align}
$P_2=5$
\begin{align}
&\widehat{Z}_0 = q^{1+\frac{53}{105}} (1-q^8-q^{16}+q^{20}-q^{24}-2 q^{28}-2 q^{32}-2 q^{44}+q^{50}+q^{52}-q^{56} + \ldots )	\,	,		\\	
&\widehat{Z}_1 = -q^{1+\frac{1}{210}} (1+q^3-q^4-q^6-q^7-q^9+2 q^{10}+q^{11}-q^{15}+q^{16}-q^{17} + \ldots )		\,	,	
\end{align}
$P_3=7$
\begin{align}
&\hspace{-3mm} \widehat{Z}_0 = q^{2+\frac{8}{105}} (1+q^{12}-q^{14}-q^{20}-q^{24}-2 q^{28}-2 q^{30}-2 q^{36}-q^{40}+2 q^{56} + \ldots )	\,	,		\\	
&\hspace{-3mm} \widehat{Z}_1 = -q^{4+\frac{121}{210}} (1+q-q^2+q^3+q^4-q^5-q^6-q^8+q^{10}-q^{11}-q^{12}+q^{13} + \ldots )	\,	.		
\end{align}

\vspace{3mm}

\subsubsection*{Analytically continued $\mathcal{R}$}

For an analytically continued $\mathcal{R}$, the character $\text{ch}_R(e^\beta)$ is not a polynomial, so we consider an expansion of $\prod_{f=1}^{F} \prod_{i<j}^{N} 2 \sinh \frac{1}{2 P_f} (\beta_i - \beta_j) /\prod_{i<j}^{N} \big(2 \sinh \frac{1}{2} (\beta_i - \beta_j) \big)^{F-1}$ in \eqref{sunint1} where the denominator of the character is taken into account.

After similar calculations as in the case of an integer $\mathcal{R}$, the integral \eqref{sunint1} becomes
\begin{align}
\begin{split}
\hspace{-5mm} \sum_{{\bf t} \in \Lambda_{\text{rt}} / H \Lambda_{\text{rt}}}
\int_{\gamma^{N-1}} \, \prod_{j=1}^{N-1} d\beta _j \,
&e^{-\frac{k}{4 \pi i} \frac{H}{P} \sum _{j=1}^{N} \beta _j^2 - k \sum _{i=1}^{N} t_j \beta _j } \, \Big( \text{ch}_R(e^\beta) \, \prod_{i<j}^{N} 2 \sinh \frac{1}{2} (\beta_i - \beta_j) \Big)	\\
&\hspace{-10mm}\times \sum_{\substack{n_{i,j}=0 \\ 1 \leq i < j \leq N}}^{\infty} \Big( \prod_{1\leq i<j \leq N} \mathcal{B}_{2P}(n_{i,j}) \Big)	\,
e^{\frac{1}{2P} \sum _{j=1}^{N} c_j \beta _j } 	\,	.
\end{split}	\label{suninta}
\end{align}
Suppose that we consider the representation $\mathcal{R}$ given by partitions $\{R_1 ; \ldots ; R_{N-1} \}$.
These $R_i$'s appear in the exponent of $e^{\beta_j}$'s, and we denote monomials in $\text{ch}_R(e^{\beta}) \prod_{i<j}^{N} \prod_{i<j}^{N} 2 \sinh \frac{1}{2} (\beta_i - \beta_j)$\footnote{
For example, for a partition $\{R_1;R_2\}$ for $G=SU(3)$, $\text{ch}_R(e^{\beta}) \prod_{i<j}^{N} \prod_{i<j}^{N} 2 \sinh \frac{1}{2} (\beta_i - \beta_j)$ is 
$-e^{R_2 \beta _1+\left(1+R_1\right) \beta _2-\beta _3}+e^{\left(1+R_1\right) \beta _1+R_2 \beta_2-\beta _3}+e^{R_2 \beta _1-\beta _2+\left(1+R_1\right) \beta _3}-e^{-\beta_1+R_2 \beta_2+\left(1+R_1\right) \beta _3}-e^{\left(1+R_1\right) \beta _1-\beta _2+R_2 \beta _3}+e^{-\beta_1+\left(1+R_1\right) \beta _2+R_2 \beta _3}$.
} as $h_{\vec{v}} \prod_{j=1}^N e^{\beta_j(L_j+ v_j)}$ where $L_j$ is given by one of $R_i$'s.
$v_j$'s are integers or half integers and $h_{\vec{v}}$ is $\pm1$.
Then, we analytically continue $R_j$ and denote
\begin{align}
u_j := \frac{R_j}{k}	\,	,
\end{align}
where we have $u_N=0$.
As before, we can directly evaluate \eqref{suninta}.
Or we can use \eqref{intrel4} and \eqref{intrelunr} for the monomial of the type $\prod_{j=1}^N e^{\beta_j(L_j+ v_j)}$ in $\text{ch}_R(e^{\beta}) \prod_{i<j}^{N} \prod_{i<j}^{N} 2 \sinh \frac{1}{2} (\beta_i - \beta_j)$, then we get
\begin{align}
\begin{split}
&e^{\pi i k \frac{P}{H} \sum_{j=1}^N t_j^2 + u_j^2 -2 t_j u'_j} \int_{\Gamma} \prod_{j=1}^{N-1} d\beta_j \, e^{-\frac{k}{\pi i} HP \sum_{j=1}^N \big( \beta_j + \pi i \frac{t_j}{H} - \pi i \frac{u'_j}{H}\big)^2 } 	\\
&\hspace{60mm}\times  \sum_{\substack{n_{i,j}=0 \\ 1 \leq i < j \leq N}}^{\infty} \Big( \prod_{1\leq i<j \leq N} \mathcal{B}_{2P}(n_{i,j}) \Big) \Big( \prod_{j=1}^{N} e^{(c_j + 2P v_j )\beta_j} \Big)	\\
&\hspace{0mm}\simeq 
e^{\pi i k \frac{P}{H} \sum_{j=1}^N t_j^2 + u_j^2 -2 t_j u'_j} 
\oint_{|z|=1} \prod_{j=1}^{N-1} \frac{dz_j}{2\pi i z_j} \sum_{\vec{m} \in \mathbb{Z}^{N}} q^{\frac{1}{8NHP} \big( N \sum_{j=1}^{N} m_j^2 -(\sum_{j=1}^{N} m_j)^2 \big) } \prod_{j=1}^{N} z_j^{m_j}	\\
&\hspace{40mm} \times e^{\frac{\pi i}{H} \sum_{j=1}^{N} (u'_j-t_j) m_j} 
\sum_{\substack{n_{i,j}=0 \\ 1 \leq i < j \leq N}}^{\infty} \Big(  \prod_{1\leq i<j \leq N} \mathcal{B}_{2P}(n_{i,j}) \Big) 
\Big( \prod_{j=1}^{N} z_j^{-(c_j+2Pv_j)} \Big)	\,	,	\label{intrelunrkn}
\end{split}
\end{align}
up to an overall factor, where $z_N = (\prod_{j=1}^{N-1}z_j)^{-1}$ and $u'_j:= \frac{L_j}{k}$, which is given by one of $u_i$'s.
As in the case of $G=SU(2)$, there are similar couplings between $t$ and $u$, and this prevents to have the usual structure with the $S$-transform when $H \geq 2$.
But when $H=1$, the $S$-transform is trivial and we obtain a $(x,q)$-series with integer coefficients,
\begin{align}
\begin{split}
\sum_{\text{monomials}} h_{\vec{v}} \, e^{\pi i k \frac{P}{H} \sum_{j=1}^N u_j^2 } \sum_{\substack{n_{i,j}=0 \\ 1 \leq i < j \leq N}}^{\infty} \Big( \prod_{1\leq i<j \leq N} \mathcal{B}_{2P}(n_{i,j}) \Big) q^{\frac{1}{8NP} \big( N \sum_{j=1}^{N} (c_j+2Pv_j)^2 -(\sum_{j=1}^{N} 2P v_j)^2 \big) }	\prod_{j=1}^N {x'_j}^{\frac{1}{2}(c_j+2Pv_j)}
\end{split}	\label{hmkn1}
\end{align}
where $x_j = e^{2\pi i u_j}$, $x_N =1$, and $x'_j = e^{2\pi i u'_j}$ is given by one of $x_j$'s depending on the monomial. 
We provide examples for the totally symmetric representation.

\subsubsection*{Case $H=1$, and totally symmetric representation}

For $G=SU(3)$, the character of the totally symmetric representation $\mathcal{S}^r$ is given by 
\begin{align}
\text{ch}_{\mathcal{R}}(e^{y}) = \frac{e^{(r+1)y_1-y_3} - e^{-y_1+(r+1)y_3} + e^{-y_1+(r+1)y_2} - e^{(r+1)y_1-y_2} + e^{-y_2+(r+1)y_3} - e^{(r+1)y_2-y_3}}{(e^{(y_1-y_2)/2}-e^{-(y_1-y_2)/2}) (e^{(y_1-y_3)/2}-e^{-(y_1-y_3)/2}) (e^{(y_2-y_3)/2}-e^{-(y_2-y_3)/2})}
\end{align}
Therefore, for Seifert integer homology sphere, $H=1$, from \eqref{hmkn1} we obtain the homological block
\begin{align}
\begin{split}
&\hspace{-3mm} F(M_3,K; x,q) =  (xq)^{P}  \sum_{\substack{n_{i,j}=0 \\ 1 \leq i < j \leq 3}}^{\infty} \prod_{1\leq i<j \leq 3} \mathcal{B}_{2P}(n_{i,j}) \,
q^{\frac{1}{8P} \sum_{j=1}^3 c_j^2}	\\
&\times \big( x^{\frac{c_1}{2}} q^{\frac{1}{2}(c_1 - c_3)} - x^{\frac{c_3}{2}} q^{-\frac{1}{2}(c_1 - c_3)} + x^{\frac{c_2}{2}} q^{-\frac{1}{2}(c_1 - c_2)} - x^{\frac{c_1}{2}} q^{\frac{1}{2}(c_1 - c_2)}  + x^{\frac{c_3}{2}} q^{-\frac{1}{2}(c_2 - c_3)} - x^{\frac{c_2}{2}} q^{\frac{1}{2}(c_2 - c_3)} \big)
\end{split}
\label{su3h1}
\end{align}
where\footnote{There is another factor $e^{\frac{k}{2\pi i}\frac{P}{3} (\log x)^2}$ in \eqref{su3h1} and similarly $e^{\frac{k}{2\pi i}\frac{3P}{8H} (\log x)^2}$ in \eqref{su4h1} from \eqref{hmkn1}, but we expect that it is cancelled out if considering the overall normalization.
}
\begin{align}
c_1 = n_{1,2} - n_{1,3}	\,	,	\quad	c_2 = -n_{1,2} - n_{2,3}	\,	,	\quad	c_3 = n_{1,3} + n_{2,3}	\,	.
\end{align}

When $G=SU(4)$, the character of the totally symmetric representation $\mathcal{S}^r$ is
\begin{align}
\begin{split}
\text{ch}_{\mathcal{R}}(e^y) &= \big(e^{\frac{1}{2} \left((2 r+3) y_1+y_2-y_3-3 y_4\right)}-e^{\frac{1}{2} \left((2 r+3) y_2+y_1-y_3-3 y_4\right)}-e^{\frac{1}{2} \left((2 r+3) y_1-y_2+y_3-3 y_4\right)}	\\
&+e^{\frac{1}{2} \left((2 r+3) y_2-y_1+y_3-3 y_4\right)}+e^{\frac{1}{2} \left(2 r y_3+y_1-y_2+3 y_3-3 y_4\right)}-e^{\frac{1}{2} \left(2 r y_3-y_1+y_2+3 y_3-3 y_4\right)}	\\
&-e^{\frac{1}{2} \left((2 r+3) y_1+y_2-3 y_3-y_4\right)}+e^{\frac{1}{2} \left((2 r+3) y_2+y_1-3 y_3-y_4\right)}+e^{\frac{1}{2} \left((2 r+3) y_1-y_2-3 y_3+y_4\right)}	\\
&-e^{\frac{1}{2} \left((2 r+3) y_2-y_1-3 y_3+y_4\right)}-e^{\frac{1}{2} \left(2 r y_4+y_1-y_2-3 y_3+3 y_4\right)}+e^{\frac{1}{2} \left(2 r y_4-y_1+y_2-3 y_3+3 y_4\right)}	\\
&+e^{\frac{1}{2} \left((2 r+3) y_1-3 y_2+y_3-y_4\right)}-e^{\frac{1}{2} \left(2 r y_3+y_1-3 y_2+3 y_3-y_4\right)}-e^{\frac{1}{2} \left((2 r+3) y_1-3 y_2-y_3+y_4\right)}	\\
&+e^{\frac{1}{2} \left(2 r y_3-y_1-3 y_2+3 y_3+y_4\right)}+e^{\frac{1}{2} \left(2 r y_4+y_1-3 y_2-y_3+3 y_4\right)}-e^{\frac{1}{2} \left(2 r y_4-y_1-3 y_2+y_3+3 y_4\right)}	\\
&-e^{\frac{1}{2} \left((2 r+3) y_2-3 y_1+y_3-y_4\right)}+e^{\frac{1}{2} \left(2 r y_3-3 y_1+y_2+3 y_3-y_4\right)}+e^{\frac{1}{2} \left((2 r+3) y_2-3 y_1-y_3+y_4\right)}	\\
&-e^{\frac{1}{2} \left(2 r y_3-3 y_1-y_2+3 y_3+y_4\right)}-e^{\frac{1}{2} \left(2 r y_4-3 y_1+y_2-y_3+3 y_4\right)}+e^{\frac{1}{2} \left(2 r y_4-3 y_1-y_2+y_3+3 y_4\right)} \big)/	\\
&((e^{(y_1-y_2)/2}-e^{-(y_1-y_2)/2}) (e^{(y_1-y_3)/2}-e^{-(y_1-y_3)/2}) (e^{(y_1-y_4)/2}-e^{-(y_1-y_4)/2}) 	\\
&(e^{(y_2-y_3)/2}-e^{-(y_2-y_3)/2}) (e^{(y_2-y_4)/2}-e^{-(y_2-y_4)/2}) (e^{(y_3-y_4)/2}-e^{-(y_3-y_4)/2}) )
\end{split}
\end{align}
and the homological block is given by
\begin{align}
\begin{split}
&\hspace{-3mm} F(M_3,K;x,q) = (x^{\frac{3}{2}} q^{\frac{5}{2}})^P \sum_{\substack{n_{i,j}=0 \\ 1 \leq i < j \leq 4}}^{\infty} \prod_{1\leq i<j \leq 4} \mathcal{B}_{2P}(n_{i,j}) \, q^{\frac{1}{8P} \sum_{j=1}^4 c_j^2} 	\\
\times &\big(
x^{\frac{c_1}{2}} q^{\frac{1}{4} \left(3c_1 +c_2-c_3-3 c_4\right)} - x^{\frac{c_2}{2}} q^{\frac{1}{4} \left(c_1+3c_2-c_3-3 c_4\right)}-x^{\frac{c_1}{2}}q^{\frac{1}{4} \left(3c_1-c_2+c_3-3 c_4\right)}	\\
&+x^{\frac{c_2}{2}}q^{\frac{1}{4} \left(-c_1+3c_2+c_3-3 c_4\right)}+x^{\frac{c_3}{2}} q^{\frac{1}{4} \left(c_1-c_2+3 c_3-3 c_4\right)}-x^{\frac{c_3}{2}} q^{\frac{1}{4} \left(-c_1+c_2+3 c_3-3 c_4\right)}	\\
&-x^{\frac{c_1}{2}}q^{\frac{1}{4} \left(3c_1+c_2-3 c_3-c_4\right)}+x^{\frac{c_2}{2}}q^{\frac{1}{4} \left(c_1+3c_2-3 c_3-c_4\right)}+x^{\frac{c_1}{2}}q^{\frac{1}{4} \left(3c_1-c_2-3 c_3+c_4\right)}	\\
&-x^{\frac{c_2}{2}}q^{\frac{1}{4} \left(-c_1+3c_2-3 c_3+c_4\right)}-x^{\frac{c_4}{2}}q^{\frac{1}{4} \left(c_1-c_2-3 c_3+3 c_4\right)}+x^{\frac{c_4}{2}}q^{\frac{1}{4} \left(-c_1+c_2-3 c_3+3 c_4\right)}	\\
&+x^{\frac{c_1}{2}}q^{\frac{1}{4} \left(3c_1 -3 c_2+c_3-c_4\right)}-x^{\frac{c_3}{2}}q^{\frac{1}{4} \left(c_1-3 c_2+3 c_3-c_4\right)}-x^{\frac{c_1}{2}}q^{\frac{1}{4} \left(3c_1 -3 c_2-c_3+c_4\right)}	\\
&+x^{\frac{c_3}{2}}q^{\frac{1}{4} \left(-c_1-3 c_2+3 c_3+c_4\right)}+x^{\frac{c_4}{2}}q^{\frac{1}{4} \left(c_1-3 c_2-c_3+3 c_4\right)}-x^{\frac{c_4}{2}}q^{\frac{1}{4} \left(-c_1-3 c_2+c_3+3 c_4\right)}	\\
&-x^{\frac{c_2}{2}}q^{\frac{1}{4} \left(-3 c_1+3c_2+c_3-c_4\right)}+x^{\frac{c_3}{2}}q^{\frac{1}{4} \left(-3 c_1+c_2+3 c_3-c_4\right)}+x^{\frac{c_2}{2}}q^{\frac{1}{4} \left(-3 c_1+3c_2-c_3+c_4\right)}	\\
&-x^{\frac{c_3}{2}} q^{\frac{1}{4} \left(-3 c_1-c_2+3 c_3+c_4\right)}-x^{\frac{c_4}{2}}q^{\frac{1}{4} \left(-3 c_1+c_2-c_3+3 c_4\right)}+x^{\frac{c_4}{2}}q^{\frac{1}{4} \left(-3 c_1-c_2+c_3+3 c_4\right)}
\big)
\end{split}
\label{su4h1}
\end{align}
where
\begin{align}
c_1 = n_{1,2} +n_{1,3} -n_{1,4}	\,	,	\	
c_2 = -n_{1,2} +n_{2,3} -n_{2,4}	\,	,	\
c_3 = -n_{1,3} -n_{2,3} -n_{3,4}	\,	,	\	
c_4 = n_{1,4} +n_{2,4} +n_{3,4}	\,	.	
\end{align}


\acknowledgments{I would like to thank the Korea Institute for Advanced Study (KIAS) for hospitality at the final stage of this work.
This work was supported by the research grant of Jeju National University in 2022.
}


\begin{appendices}

\section{Lens space and representations of $G=SU(2)$}
\label{sec:lens}

In this appendix, we discuss the calculation in section \ref{sec:su2} for the case of the lens space when $G=SU(2)$ and the result of \cite{Gukov-Pei-Putrov-Vafa}.

When $G=SU(2)$, the contribution from abelian flat connections for a knot in the lens space is given by
\begin{align}
\sum_{t=0}^{p-1} e^{\frac{2 \pi i}{p}kt^2 } 
\int_\gamma dy e^{-\frac{k}{2\pi i} p \big(y+ \frac{2\pi i}{p}t \big)^2} \chi_R(y) (e^y - e^{-y})^2	\,	.	\label{lens}
\end{align}
By using \eqref{intrel1}, \eqref{lens} can be expressed as
\begin{align}
\pi \bigg( \frac{2i}{k} \frac{1}{p} \bigg)^{1/2} \sum_{t=0}^{p-1} e^{\frac{2\pi i}{p}kt^2} \oint \frac{dz}{2 \pi i z} \sum_{m \in \mathbb{Z}} q^{\frac{m^2}{4p}} z^m e^{2\pi i \frac{tm}{p}} \chi_R(z) (z-z^{-1})^2	\,	.	\label{lens2}
\end{align}
We can express the sum $\sum_{m \in \mathbb{Z}}$ as $\sum_{b \in \mathbb{Z} / 2p \mathbb{Z}} \sum_{l \in 2p \mathbb{Z} +b}$, then \eqref{lens2} can be put in the form of
\begin{align}
\pi \bigg( \frac{2i}{k} \frac{1}{p} \bigg)^{1/2} \sum_{t=0}^{p-1} e^{\frac{2\pi i}{p}kt^2} \sum_{b \in \mathbb{Z}/2p \mathbb{Z}} e^{\frac{2\pi i }{p} tb} \oint \frac{dz}{2 \pi i z} \chi_R(z) (z-z^{-1})^2 \sum_{l \in 2p\mathbb{Z} + b} q^{\frac{l^2}{4p}} z^l	\,	. 	\label{lens3}
\end{align}
We may instead use $\sum_{b \in \mathbb{Z} / p \mathbb{Z}} \sum_{l \in p \mathbb{Z} +b}$.
However, since the denominator of $q^{\frac{l^2}{4p}}$ is $4p$ and exponents of $z$ in $\chi_R(z) (z-z^{-1})^2$ are all odd or all even depending on $R$, it is more natural to have an expression as in \eqref{lens3}.
From \eqref{lens3}, we see that the homological block is given by
\begin{align}
\widehat{Z}_b(M_3,K;R,q) = \frac{1}{|W_b|} \oint \frac{dz}{2 \pi i z} \chi_R(z) (z-z^{-1})^2 \Big(\sum_{l \in 2p\mathbb{Z} + b} + \sum_{l \in 2p\mathbb{Z} - b} \Big) q^{\frac{l^2}{4p}} z^l 	\label{lenshb}
\end{align}
Suppose that $R$ is odd, then exponents of $z$ in $\chi_R(z) (z-z^{-1})^2$ are all even, so $b$ should also be even for non-zero results.
We denote such $b$ as $b= 2 \sf{b}$, then \eqref{lens3} is given by
\begin{align}
\pi \bigg( \frac{2i}{k} \frac{1}{p} \bigg)^{1/2} \sum_{t=0}^{p-1} e^{\frac{2\pi i}{p}kt^2} \sum_{\sf{b} =0}^{p-1} e^{\frac{4\pi i }{p} t \sf{b}} \oint \frac{dz}{2 \pi i z} \chi_R(z) (z-z^{-1})^2 \sum_{l \in p\mathbb{Z} + \sf{b}} q^{\frac{l^2}{p}} z^{2l}	\,	. 	\label{lense}
\end{align}
This is the expression that is obtained in \cite{Gukov-Pei-Putrov-Vafa}.

Meanwhile, \eqref{lens3} gives non-zero homological blocks also when $R$ is even.
For this case, exponents of $z$ in $\chi_R(z) (z-z^{-1})^2$ are all odd, we have non-zero results when $b$ is odd.
For odd $p$, we obtain the homological block \eqref{lenshb} with the usual $S$-transform.\footnote{In this case, $b=p$ plays a role of the label $``0"$ of $\widehat{Z}_0$.} 
However, for even $p$, we can have homological blocks but we don't have the $S$-transform whose square is an identity matrix.\footnote{This is not like the case, for example, when $G=SU(2)$ and $M_3$ is a Seifert manifold with three singular fibers with $H$ being 2 mod 4 where the ``$S$-transform" can be expressed as a tensor product of matrices that contains a matrix whose square is the identity.
}
This unusual ``$S$-transform" when $p$ is even can be seen from that there is no $b$ that can play a role of the label $``0"$ of $\widehat{Z}_0$ in $e^{\frac{2\pi i}{p} tb}+e^{-\frac{2\pi i}{p} tb}$.

We also note that existence of homological blocks for odd $b$ is consistent with the calculation of the index of $T[M_3]$ on $S^2 \times S^1$ in \cite{Gukov-Pei-Putrov-Vafa}.
This is because for the lens space it was checked that the index with a loop located at a pole of $S^2$ can be expressed as factorization where one of the factors is the homological block without a knot and another is the homological block for a knot.
When there is no knot, $b$ is even, so the product with the homological blocks for a knot that are labelled by odd $b$ gives the zero index automatically, which is indeed the case for the $S^2 \times S^1$ index when the $R$ is even.

\section{Remarks on a knot complement in Seifert rational homology spheres}
\label{sec:rmk}

We saw in section \ref{ssec:su2rshs} that for a knot complement in Seifert rational homology spheres with the analytically continued $\mathcal{R}$ the structure that is obtained for Seifert knots with integer $\mathcal{R}$ or Seifert manifolds without a knot doesn't arise.
At the level of equations, this is due to the factor $e^{\pm 2\pi i k \frac{P}{H} t u}$ in \eqref{su2inthr1}, \textit{i.e.} the coupling between $u$ and $t$.
This term also arises when $R$ is an integer as seen in \eqref{su2inth33}.
However, \eqref{su2inth33} can be arranged to be equal to \eqref{su2inth31} when $R$ is an integer.

In section \ref{ssec:su2rshs}, we have set $R$ to be analytically continued on the right-hand side of \eqref{su2inthr1} and then took the limit $q \searrow e^{\frac{2\pi i}{k}}$ and $x \rightarrow e^{2\pi i \frac{R}{k}}$.
In this equation, one may just take $R$ to be an integer at $e^{2\pi i \frac{P}{H}tR}$, but still analytically continued at $x$ which is eventually taken to the limit $x \rightarrow e^{2\pi i \frac{R}{k}}$ in order to produce the Wilson loop expectation value with integer $k$ and $R$.
This leads, for example when $H$ is odd, to
\begin{align}
\begin{split}
&\hspace{-3mm} e^{\frac{\pi i}{2} k \frac{P}{H} u^2} \Bigg( \sum_{b \in \mathbb{Z}/2H\mathbb{Z}} 
\oint \frac{dz}{2\pi i z} \sum_{l \in 2H\mathbb{Z}+b} q^{\frac{l^2}{4HP}} z^{-l} (x^{\frac{l}{2H}} - x^{-\frac{l}{2H}}) \sum_{n=0}^{\infty} \mathcal{B}_{2P}(n) z^{n}	\\
&\hspace{-4mm} +e^{2\pi i k \frac{P}{H}t^2} \sum_{b \in \mathbb{Z}/2H\mathbb{Z}}  (e^{2 \pi i \frac{t}{H}(b+PR)} + e^{-2 \pi i \frac{t}{H}(b+PR)})
\oint \frac{dz}{2\pi i z} \sum_{l \in 2H\mathbb{Z}+b} q^{\frac{l^2}{4HP}} z^{-l} \sum_{n=0}^{\infty} \mathcal{B}_{2P}(n) z^{n} x^{\frac{l}{2H}}	\\
&\hspace{-4mm} -e^{2\pi i k \frac{P}{H}t^2} \sum_{b \in \mathbb{Z}/2H\mathbb{Z}}  (e^{2 \pi i \frac{t}{H}(b-PR)} + e^{-2 \pi i \frac{t}{H}(b-PR)})
\oint \frac{dz}{2\pi i z} \sum_{l \in 2H\mathbb{Z}+b} q^{\frac{l^2}{4HP}} z^{-l} \sum_{n=0}^{\infty} \mathcal{B}_{2P}(n) z^{n} x^{-\frac{l}{2H}} \Bigg) \Bigg|_{q \searrow e^{\frac{2\pi i }{k}}, \, x\rightarrow e^{2\pi i \frac{R}{k}}}	\,	.
\end{split}	\label{su2inthr3}
\end{align}
where the second and the third line of \eqref{su2inthr3}, \textit{i.e.} the part for $t \neq 0$, can be further organized to
\begin{align}
\begin{split}
&(e^{2 \pi i \frac{t}{H}PR} + e^{-2 \pi i \frac{t}{H}PR})
\oint \frac{dz}{2\pi i z} \sum_{l \in 2H\mathbb{Z}+H} q^{\frac{l^2}{4HP}} z^{-l} (x^{\frac{l}{2H}} - x^{-\frac{l}{2H}}) \sum_{n=0}^{\infty} \mathcal{B}_{2P}(n) z^{n} 	\\
&+\sum_{b=\{1,3, \ldots, H-2 \}} (e^{2 \pi i \frac{t}{H}(b+PR)} + e^{-2 \pi i \frac{t}{H}(b+PR)})	\\
&\hspace{10mm}\times
\oint \frac{dz}{2\pi i z} \Big( \sum_{l \in 2H\mathbb{Z}+b} q^{\frac{l^2}{4HP}} z^{-l} x^{\frac{l}{2H}} - \sum_{l \in 2H\mathbb{Z}-b} q^{\frac{l^2}{4HP}} z^{-l} x^{-\frac{l}{2H}}\Big) \sum_{n=0}^{\infty} \mathcal{B}_{2P}(n) z^{n}	\\
&+\sum_{b=\{1,3, \ldots, H-2 \}} (e^{2 \pi i \frac{t}{H}(b-PR)} + e^{-2 \pi i \frac{t}{H}(b-PR)})	\\
&\hspace{10mm}\times \oint \frac{dz}{2\pi i z} \Big( \sum_{l \in 2H\mathbb{Z}-b} q^{\frac{l^2}{4HP}} z^{-l} x^{\frac{l}{2H}} - \sum_{l \in 2H\mathbb{Z}+b} q^{\frac{l^2}{4HP}} z^{-l} x^{-\frac{l}{2H}} \Big) \sum_{n=0}^{\infty} \mathcal{B}_{2P}(n) z^{n} 
\end{split}	\label{su2inthr4}
\end{align}
up to an overall factor.
Each integral part in \eqref{su2inthr3} and \eqref{su2inthr4} with a given $b$ gives $(x,q)$-series with integer coefficients, but in general it is still not possible to obtain the structure with the $S$-transform that arises in the case of an integer $R$ or closed 3-manifolds.

Hence, from the discussion above, it is unlikely the case that there are homological blocks with a proper $S$-transform when $M_3$ is a Seifert rational homology sphere and $\mathcal{R}$ is analytically continued.
It might be possible that this is due to a limitation of the approach discussed in this paper, but it would not be the case since the integral expression with the Gaussian factor, which \eqref{su2inthr1} is obtained from, is an appropriate quantity, considering the resurgent analysis.	\\

For a knot complement in $S^3$, we have a single homological block $F(S^3,K; x,q)$, and it is consistent with the interpretation as a half index of $T[S^3 \backslash K]$ with a boundary condition.
However, from the above analysis, there seems no homological blocks with a proper $S$-transform for an analytically continued $\mathcal{R}$ in a Seifert rational homology sphere $M_3$.
So the interpretation in the context of the 3d-3d correspondence is unclear because the half index of the corresponding $T[M_3\backslash K]$ with proper boundary conditions would give $(x,q)$-series with integer coefficients.
It would be interesting to understand this issue better.

\end{appendices}

\newpage

\bibliographystyle{JHEP}
\bibliography{ref}

\end{document}